\documentclass[fleqn,usenatbib]{mnras}
\usepackage[T1]{fontenc}
\usepackage{aecompl}

\usepackage[utf8]{inputenc}
\usepackage{bm}
\usepackage{caption}

\usepackage[dvipsnames]{xcolor}

\usepackage{graphicx}	
\usepackage{amsmath}	
\usepackage{amssymb}	
\usepackage{xcolor}
\usepackage{ulem}
\usepackage{rotating}
\usepackage{orcidlink}
\usepackage{xspace}
\usepackage{nicefrac}
\usepackage{orcidlink}

\usepackage{pifont}

\newcommand{\Myr}{~\text{Myr}}

\newcommand{\Msun}{~\text{M}_\odot}
\newcommand{\pc}{~\text{pc}}
\newcommand{\kpc}{~\text{kpc}}

\newcommand{\cMpc}{~\text{cMpc}}

\newcommand{\percent}{~\text{per~cent}}
\newcommand{\AAA}{{\text{\AA}}}
\newcommand{\bh}{{\rm BH}}

\newcommand{\effrad}{{\epsilon_{\rm r}}}

\newcommand{\cmark}{\ding{51}}%

\newcommand{\simname}{\textsc{meli$\odot$ra}\xspace}
\newcommand{\ramses}{\textsc{ramses}\xspace}

\newcommand{\boxsizeL}{{5}} 
\newcommand{\finalz}{{4}} 
\newcommand{\levelmin}{{8}} 
\newcommand{\levelmax}{{16}} 
\newcommand{\spatialresmin}{{76}} 
\newcommand{\spatialresmax}{{20}} 
\newcommand{\nsf}{{1}} 

\newcommand{\NEW}[1]{#1}

\usepackage[dvipsnames]{xcolor}
\definecolor{M000_color}{HTML}{66cdaa}
\definecolor{M002_color}{HTML}{9467bd}
\definecolor{M004_color}{HTML}{4169e1}

\title[LRDs as DCBH nurseries]{Little Red Dots as Direct-collapse Black Hole Nurseries}

\author[Cenci et al.]{
\parbox{\textwidth}{
Elia Cenci$^{\,\dagger}$\orcidlink{0000-0002-0766-1704}\thanks{E-mail:\href{mailto:elia.cenci@unige.ch}{elia.cenci@unige.ch}},
Melanie Habouzit$^{\,\dagger}$\orcidlink{0000-0003-4750-0187}
\vspace{6pt}
}\\
$^{\dagger}$Department of Astronomy, University of Geneva, Chemin Pegasi 51, Versoix CH-1290, Switzerland \\
}

\date{Accepted XXX. Received YYY; in original form ZZZ}

\pubyear{2024}

\begin{document}
\label{firstpage}
\pagerange{\pageref{firstpage}--\pageref{lastpage}}
\maketitle

\begin{abstract}
The James Webb Space Telescope recently uncovered a population of massive black holes (BHs) in the first billion years after the Big Bang. Among these high-redshift BH candidates, observations have identified a class of active galactic nuclei candidates, dubbed Little Red Dots (LRDs), with extraordinarily compact gas reservoirs and peculiar spectral features. LRDs clearly emerge at redshift $z\lesssim 8$ and their abundance declines by $z\lesssim 5$. Recent theoretical studies have explored the link between LRDs and the formation of heavy BH seeds in the early Universe, such as direct-collapse BHs (DCBHs). Here we present results from preliminary runs for the \simname cosmological hydrodynamical simulations, where we implement an accurate model for DCBH formation, accounting for the Lyman-Werner radiation field and mass-inflow rates in the target host haloes. We aim to test whether or not DCBH formation could lead to systems resembling those hypothesized for LRDs. We find that the population of newly formed DCBHs in the simulations exhibits a steep decline at $z\lesssim 6$, akin to the emergence of LRDs, primarily driven by reduced inflows. The birth of DCBHs is associated with a significant gas compaction event, followed by a phase of intense luminosity in the $200\Myr$ after their birth, and subsequently by the formation of the first PopIII stars in these very haloes. If these DCBHs nurseries are associated with LRDs, then it could explain their weak emission from X-rays and hot dust.
\end{abstract}

\begin{keywords}
methods: numerical -- galaxies: formation -- galaxies: active -- quasars: supermassive black holes
\end{keywords}

\section{Introduction}\label{sec:introduction}
\begin{figure*}
    \centering
    \includegraphics[width=\hsize]{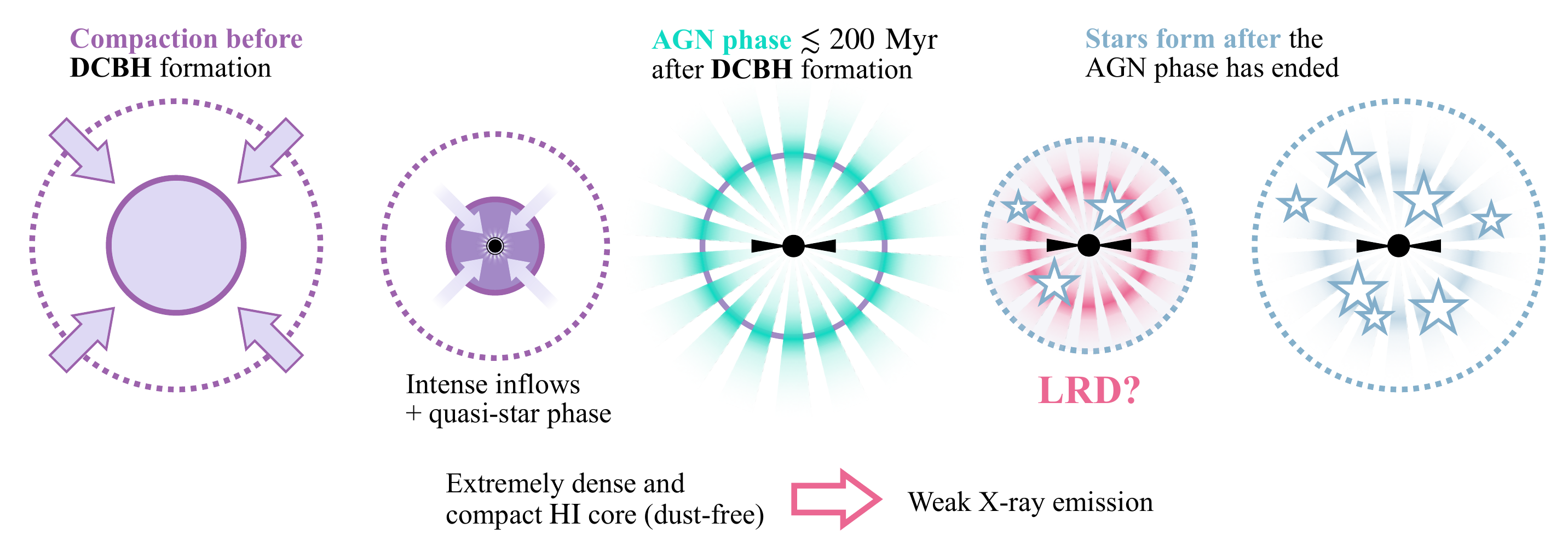}
    \caption{Schematic illustration of the scenario that we explore with this work. A gas compaction event, associated with increasing gas density and shrinking of effective size of the gas reservoir, funnels material towards the central regions of atomic-cooling haloes. After a intermediate phase of super-massive star, a DCBH forms. Intense accretion onto the newborn DCBH results in a short-lived AGN phase, with a high bolometric luminosity. After $\lesssim 200\Myr$, the first PopIII stars start to form and the AGN luminosity decreases. In the short period of time when we can observe the emission from both the AGN and PopIII stars, this system could resemble a LRD. As gas increases in size and becomes more diffuse, and inflows become less efficient, the AGN phase ends.}
    \label{fig:cartoon}
\end{figure*}

In the recent years, observations with the James Webb Space Telescope (JWST) have revealed a population of faint, broad-line active galactic nuclei (AGN) at $z>5$ \citep[][]{Onoue2023,Kocevski2023,Harikane2023,Akins2024,Barro2024,Furtak2024,Greene2024,Killi2024,Leung2024,Maiolino2024,Matthee2024a}, powered by BHs with masses of about $10^6-10^7\Msun$. A significant fraction of these AGN have compact morphologies \citep[with effective radii $\lesssim 100-300\pc$; e.g,][]{Furtak2023,Baggen2024,Casey2024,Labbe2025} and red rest-frame optical colours. 
These sources have been dubbed Little Red Dots \citep[LRDs;][]{Matthee2024a}, and they are characterised by peculiar spectral energy distributions (SEDs). Specifically, they have V-shaped SEDs, with a red slope in the rest-frame optical and a blue slope in the rest-frame ultraviolet (UV). The population of LRDs outnumbers that of $z<9$ quasars by 1 dex, making the practically ubiquitous in the early Universe, but it dramatically declines after $z\lesssim 6$ \citep[e.g.,][]{Kocevski2024}. The low-redshift decrease in the LRDs reported abundance is not associated with selection effects, contrary to the possibly incomplete samples at $z>9$ \citep[see][]{Inayoshi2025}.

LRDs are thought to reside in relatively low-mass haloes ($M_{\,\rm halo}\lesssim 10^{11}\Msun$), and constitute a non-negligible fraction of the entire galaxy population \citep[about 1\percent; see, e.g.,][]{Matthee2024b,Arita2025,Pizzati2025}, with abundances of order $10^{-5}-10^{-4}~\mathrm{cMpc}^{-3}$ \citep[e.g.,][]{Harikane2023,Akins2024,Greene2024,Kokorev2024,Maiolino2024,Matthee2024a,Lin2025}.

The origin of their SEDs is still debated, with the most popular scenarios involving either dusty star formation \citep[e.g.,][]{Baggen2024,Barro2024,Williams2024} or a reddened AGN \citep[e.g.,][]{Onoue2023,Akins2024,Labbe2024,Ubler2024,Li_Z_2025,Volonteri2025}. However, LRDs may not trace an homogeneous population with a common explanation for their origin and appearance\NEW{, possibly related to varied  and non-homogeneous selection criteria \citep[e.g.][]{Hviding2025}}. The broad emission lines \citep[$1000-2000~\mathrm{km/s}$;][]{Greene2024,Kocevski2024,Taylor2024} observed in LRDs are consistent with a scenario involving an AGN \citep[or a very dense stellar cores; see][]{Baggen2024}. However, LRDs seem to lack hot dust thermal emission \citep[$\sim 1-3~\mu\mathrm{m}$; e.g.,][]{Perez-Gonzalez2024,Setton2025}, expected from a dusty torus in the vicinity of the BH, as typically observed in high-redshift AGN. Furthermore, LRDs have unexpectedly weak of X-ray detections given the expected AGN luminosities \citep[e.g.,][]{Ananna2024,Lambrides2024,Yue2024}. Their X-ray spectrum and emission lines also suggest that LRDs might be associated with high hydrogen columns densities of BHs accreting above their Eddington limit \citep[e.g.,][]{Inayoshi2024,Lambrides2024,Pacucci_and_Narayan2024,Madau2025}.
Recent theoretical work suggest that the emergence of LRDs, as well as their spectral properties, can be mostly explained as early growth phases of massive BH seeds in the early Universe \citep[see][]{Inayoshi2025,Inayoshi_and_Maiolino2025} or their super-massive stars  progenitors in the early Universe \citep{Begelman_and_Dexter2025}. The heaviest seeds (with masses of about $10^4-10^6\Msun$) are thought to form via isothermal monolithic collapse of high density gas cores in pristine, atomic-cooling halos \citep[e.g.,][]{Loeb_and_Rasio1994,Eisenstein_and_Loeb1995,Haiman1996a,Koushiappas2004,Begelman2006,Lodato_and_Natarajan2006,Latif2013a,Johnson2014,Choi2015}, after an intermediate stage of supermassive-star that accretes mass with sustained rates of $\sim 0.01-1\Msun\,\mathrm{yr}^{-1}$ for a few Myr \citep[e.g.,][]{Begelman2008,Begelman2010,Latif2013e,Prieto2013,Matsukoba2019,Nandal2023,Prole2024a}. For these direct-collapse black holes (DCBHs) to form, their host haloes must have low gas-phase metallicities \citep[$Z/Z_\odot \lesssim 10^{-5} - 10^{-4}$; e.g.,][]{Clark2008,Omukai2008,Latif2016,Chon_and_Omukai2020} and low H$_2$ fractions \citep[e.g.,][]{Bromm_and_Loeb2003}. \NEW{These} haloes are expected to experience intense UV radiation in the Lyman-Werner (LW) band\footnote{These photons are in the energy range 11.2 - 13.6~eV, that can photo-dissociate $\rm H_2$. The LW flux is usually given in units of $J_{21} = 10^{-21}$ erg s$^{-1}$ cm$^{-2}$ sr$^{-1}$ Hz$^{-1}$. Here, we will use $J_{\rm LW}$ to refer to the average intensity in the LW band. Some authors use instead the intensity at the energy of 13.6~eV.}, from nearby star-forming regions \citep[e.g.,][]{Omukai2001,Johnson_and_Bromm2007,Dijkstra2008,Shang2010,Johnson2011,Wolcott-Green2011,Agarwal2012,Safranek-Shrader2012,Latif2013c,Regan2014,Luo2020}. \NEW{Alternatively, the formation of heavy seeds could be favoured by, e.g., high relative (streaming) velocities between pristine gas and dark matter \citep[][]{Schauer2019}, rapidly assembling halos \citep[][]{Wise2019}, converging flows \citep[][]{Latif2022b}, and halo mergers \citep[][]{Prole2024b}. Furthermore, recent observations and numerical studies claim that heavy seeds could originate from primordial BHs \citep[e.g.,][]{Dayal2025,Maiolino2025b,Prole2025}.}

Numerical simulations offer a powerful framework to give predictions based on different models for the formation of early BHs and their host galaxies. Large-volume simulations
typically adopt a rather simplistic approach for BH seeds, e.g., spawning fixed mass
BHs in selected massive haloes. This is the case for simulations such as, e.g., Illustris \citep[][]{Genel2014,Vogelsberger2014a,Vogelsberger2014b,Sijacki2015} and EAGLE \citep[][]{Crain2015,Schaye2015}. However, some simulations do include criteria for DCBH formation that account for gas density, metallicity, and UV radiation field \citep[e.g.,][]{Chon2016,Tremmel2017,Chiaki2023,Regan2023b,Bhowmick2024b,Bhowmick2024c,Jeon2024}. Different simulations and semi-analytical models yield different properties of DCBHs, mainly due to differences in their modelling of either radiation or distribution of metals \citep[][]{Habouzit2016}.

In this paper, we explore the possible connection between LRDs and newly formed DCBHs, making use of a new set of cosmological hydrodynamical simulations. We present results from preliminary runs from the \simname simulations (see Cenci et al., in preparation), where we implement an accurate set of criteria and a novel model for the formation of DCBHs in a cosmological volume.

This paper is structured as follows. In Section~\ref{sec:methods}, we introduce the \simname simulations analysed in this work and summarize the physics models, including our new DCBH formation model. In Section~\ref{sec:decline} and~\ref{sec:evolution}, we discuss our results, in particular the evolution of the population of newly-formed DCBHs and the properties of their host haloes around the DCBH formation time. In Section~\ref{sec:SED}, we discuss the implications of our results in relation to the observed spectra of LRDs. Finally, in Section~\ref{sec:conclusions}, we summarize our findings and conclusions.


\section{Methods}\label{sec:methods}
In this section, we describe the relevant properties of the sub-set of the \simname simulations that we analyse in this work, and summarize the physical models we employed, especially our new model for DCBH formation in cosmological runs (see Section~\ref{sec:DCBH_model_methods}). For a more detailed description of the full simulation suite, we refer the reader to our upcoming method paper (Cenci et al., in preparation).

\subsection{Simulations}
In \simname, we simulate a cosmological volume with periodic boundary conditions, evolved to a final redshift of $z=\finalz$. For this work, we make use of preliminary runs with a volume of $\left(\boxsizeL\cMpc\right)^3$.

Initial conditions are generated at $z=100$, with the MUlti Scale Initial Conditions \citep[MUSIC;][]{Hahn_and_Abel2011} code. We assumed cosmological parameters $\Omega_{\rm m}=0.272$, $\Omega_\Lambda=1-\Omega_{\rm m} = 0.728$, $\Omega_{\rm b}=0.045$, $h=0.704$, $\sigma_8=0.81$, $n_{\rm s}=0.967$, consistent with results from the Wilkinson Microwave Anistropy Probe \citep[WMAP-7; measurements taken over 7 years of observations;][]{Komatsu2011}. We employed a \citet[][]{Eisenstein_and_Hu1998,Eisenstein_and_Hu1999} transfer function for a cold dark matter cosmology, that includes perturbations due to several baryon effects.

Gravity and hydrodynamics in \simname are solved with the adaptive mesh refinement \ramses code\footnote{The publicly available repository for \ramses can be found at: \url{https://bitbucket.org/rteyssie/ramses/src/master/}.} \citep[][]{Teyssier2002}. \ramses is fundamentally based on a fully threaded oct-tree structure \citep[e.g.,][]{Kravtsov1997,Khokhlov1998}.

The \simname simulations start from a minimum refinement level $\ell_{\rm ref, min}=\levelmin$ (coarse level), corresponding to a coarse resolution of $\Delta x_{\rm max}\simeq\spatialresmax\kpc$ and a dark matter mass resolution of $m_{\rm DM}\simeq 2\times 10^5 \Msun$. We adaptively refine the grid to a maximum refinement level $\ell_{\rm ref, max}=\levelmax$, achieving a maximum spatial resolution (i.e., the physical cell size at $\ell_{\rm ref, max}$) as high as $\Delta x_{\rm min}\simeq\spatialresmin\pc$. \NEW{Refinement is allowed over the the entirety of the simulated volume.} The criterion for mesh refinement follows a quasi-Lagrangian approach: for a given cell, refinement is triggered if
\begin{equation}
    M_{\rm DM} + \frac{\Omega_{\rm m}}{\Omega_{\rm b}}\,M_{\rm b} \,\geq\, 8\,m_{\rm DM}
    ~,
\end{equation}
\noindent where $M_{\rm DM}$ and $M_{\rm b}$ are the total dark matter and baryon (gas + stars) mass in the cell, respectively. Additionally, a cell is refined if its size is larger than twice the characteristic Jeans length computed from the cell's temperature and density. Each refinement level consists of a characteristic cell size that is half of that of the previous coarser level. The cells containing sink particles (representing, e.g., BHs) are enforced to get refined to the maximum refinement level that is currently allowed. To keep an approximately constant physical scale-length for grid cells at the finest refinement level, we allow for an additional refinement level (above $\ell_{\rm ref, min}$) every time the scale factor doubles.

Euler equations for gas dynamics on the grid are solved with a unsplit, second-order MUSCL-Hancock scheme \citep[][]{van_Leer1979} and the Harten-Lax-Van Leer-Contact \citep[HLLC][]{Toro1994,Toro1999} approximate Riemann solver. The dynamics of collisionless particles (dark matter, stars, and BHs) is solved for using a particle-mesh (PM) method with cloud-in-cell (CIC) interpolation. Time-steps of finer levels in the mesh hierarchy are updated twice for every update step in the next coarser levels. For stability, time-steps are limited following the Courant–Friedrichs–Lewy condition, with a Courant number of 0.8.

\renewcommand{\arraystretch}{1.5} 
\begin{table}
    \centering
    \caption{Summary of the DCBH formation criteria used in the sub-set of \simname simulations that are used in this work. The locally adjusted thresholds $J_{\,\mathrm{LW},\,\rm loc}^{\,\rm crit}$ and $\dot{M}^{\,\rm crit}_{\,\rm in,\rm loc}$ are given by Equations~\eqref{eqn:J_LW_crit} and ~\eqref{eqn:Mdot_crit}, respectively.}
    \label{tab:runs}
    \begin{tabular}{|p{2.3cm}|lcccc}
        \hline
        Simulation label && \textcolor{M000_color}{\textbf{M000}} & \textcolor{M002_color}{\textbf{M002}} & \textcolor{M004_color}{\textbf{M004}}\\
        \hline
        $M_{\rm star}=0$
        && \cmark & \cmark & \cmark\\
        $\bar{n}>n_{\,\rm SF}$
        && \cmark & \cmark & \cmark \\
        $\log_{10}\,Z/Z_\odot$
        && $<-4$ & $<-4$ & $<-6$ \\
        $J_{\,\rm LW}~\left[J_{21}\right]$
        && $>J_{\,\mathrm{LW},\,\rm loc}^{\,\rm crit}$ & $>100$ & $>100$ \\
        $\dot{M}_{\,\rm in}~\left[\mathrm{M}_\odot/\mathrm{yr}\right]$
        && $>\dot{M}^{\,\rm crit}_{\,\rm in,\rm loc}$ & $>0.1$ & $>1$\\
        \hline
    \end{tabular}
\end{table}

\subsection{Clump finder}
Haloes in our simulations are associated with gas density clumps, that are identified throughout the simulations with the built-in clump-finder routines of \ramses \citep[][]{Bleuler2014,Bleuler2015}. Gas cells with a mass density above the density threshold $\hat{\rho}_{\rm peak}=80\,\rho_{\rm crit,\,m}$ are assigned to the nearest density peak, following the steepest-ascent path. Here $\rho_{\rm crit,\,m}=1.88\times 10^{-29}\Omega_{\rm m}\,h^2\,\left(1+z\right)^3~\left[\mathrm{g}/\mathrm{cm^{3}}\right]$ is the critical matter density of the Universe at the given redshift. Saddle points between peaks, where density is larger than $\hat{\rho}_{\rm saddle}=200\,\rho_{\rm crit,\,m}$, are also identified in order to merge the non-relevant structures to their nearby clumps. When the peak-to-saddle density ratio is below a relevance threshold $\chi_{\rm clump}=3$, the peaks are merged together. Otherwise, both peaks survive and be considered as independent clump structures.

\subsection{Gas cooling and heating}\label{sec:gas_cooling}
The equation of state of gas in \simname is that of an ideal gas with a adiabatic index $\gamma=5/3$ (monoatomic gas). Gas cooling follows the model by \citet[][]{Sutherland_and_Dopita1993}, accounting for cooling by H, He, and metals, down to $10^4~\mathrm{K}$. At temperatures $\lesssim 10^4~\mathrm{K}$, we include fine-structure, infrared line cooling through the analytical approximation of the cooling rates by \citet[][]{Rosen_and_Bregman1995} \citep[based on the results by][]{Dalgarno_and_McCray1972}. We assume an initial average gas-phase metallicity $Z_{0}=0$. After redshift $z_{\rm reion}=8.5$, we consider gas heating from a uniform UV background from high-redshift reionization sources, following the model by \citet[][]{Haardt_and_Madau1996}. In order to mimic the effect of gas heating in star-forming regions, and prevent excessive spurious fragmentation, we impose a polytropic temperature floor $T_{\rm floor} = T_{0}\,\left(n/n_{\rm SF}\right)^{\kappa-1}$ of the high-density inter-stellar medium, where $\kappa=1.6$ and $T_{0}=10^3~\mathrm{K}$. At high densities, above the density threshold for star formation, $n_{\rm SF}$, this prescription results in an artificially increased thermal pressure that prevents gas from cooling to lower temperatures \citep[e.g.,][]{Springel_and_Hernquist2003}.

\subsection{Star formation}
Star formation in \simname takes place in gas with hydrogen density above a fixed threshold $n_{\rm SF}=\nsf~\mathrm{cm}^{-3}$. A gas cells that satisfy this density criterion form stars with an expected, average star-formation rate density $\dot{\rho}_{\star}$ following the \citet[][]{Schmidt1959} law:
\begin{equation}
    \dot{\rho}_{\star} = \epsilon_{\rm ff}\,\frac{\rho}{t_{\rm ff}}
    ~,
\end{equation}
\noindent where $\rho$ is the cell's gas mass density, $t_{\rm ff}=\sqrt{3\pi/32\,G\,\rho\;}$ is the local free-fall time, and $\epsilon_{\rm ff}$ is the \textit{local} star formation efficiency per free-fall time. While often taken as a fix parameter independent of redshift, here $\epsilon_{\rm ff}$ is estimated based on the local properties of the inter-stellar medium, following the turbulence-regulated model by \citet[][]{Federrath_and_Klessen2012} and \citet[][]{Padoan_and_Nordlund2011} \citep[see also][]{Hennebelle_and_Chabrier2011}.

Over a time-step $\Delta t$, a gas cell with a $\dot{\rho}_{\star}>0$ forms a star particle that represents a population of a number $N_\star$ of stars, that follows a Poisson statistics\NEW{, and limited such that every gas cell can convert up to maximum $90\percent$ of their available gas mass into stars}. \NEW{The minimum resolved stellar mass in the simulations used in this work is $m_{\star,0}\simeq 10^4\Msun$.} The newly formed star particles have a total mass $m_\star=N_\star\,m_{\star,0}$ and inherit momentum ($p_\star$) and metallicity ($Z_\star$) from their parent gas cell. 

Star particles with metallicity $Z/Z_\odot<10^{-4}$ are classified as PopIII stars, whereas PopII stars have larger metallicities \citep[see, e.g.,][]{Maio2011}. PopIII stars in \simname form with a Salpeter-like initial-mass function (IMF) \citep[][]{Salpeter1955,Miller_and_Scalo1979,Kimm2017}\NEW{, whereas other stars form with a \citet[][]{Kroupa2001} IMF.}

\subsection{Stellar feedback}
We collectively consider feedback from type II and type Ia supernovae (SNe), that deposit thermal energy, kinetic energy, and the metals in the surrounding medium. Furthermore, we distinguish between feedback from PopIII and PopII stars. Star particles that are older than $10\Myr$ undergo SN explosions that eject a fixed fraction $\eta_{\rm SN}=0.2$ of their mass. \NEW{This is numerically equivalent to assuming that $20\percent$ of the stars simultaneously go off as SNe in the coeval population represented by simulated star particles.} The ejected mass is instantaneously deposited in the 27 neighbouring father cells, with a metal yield of $Y_{\rm SN}=0.1$. Therefore, when triggering SN events, a star particle ejects a mass $M_{\mathrm {ej}}=\eta_{\rm SN}\,m_{\star}$, with a metallicity $Z_{\mathrm{ej}}=Z_{\star}\,+\,Y_{\rm SN}(1-Z_{\star})$. Furthermore, SNe release a total energy of $E_{\rm SN}=10^{51}~\mathrm{erg}$ per $10\Msun$ of ejected mass. \NEW{In our simulations, we do not resolve the radius of SNe remnants at the end of the Sedov–Taylor phase and the energy ($E_{\rm SN}$) that is released is instantaneously deposited as thermal energy into nearby gas. In \simname, PopIII star particles can go off as SNe after only $3\Myr$ \citep[approximately the average lifetime for a $40\Msun$ PopIII star; see][]{Schaerer2002,Schaerer2003}. We assume an ejecta mass fraction of $\eta_{\rm SN}=0.35$, with a total energy release of $10^{51}~\mathrm{erg}$ per $10\Msun$ of ejected mass and a metal yield of $Y_{\rm SN}=0.2$ \citep[see][and references therein]{Wise2012a,Kimm2017}.}

Massive young stars likely reside in dense gas that can efficiently cool, quickly radiating away the thermal energy deposited by SN feedback \citep[e.g.,][]{Katz1992}. However, non-thermal processes (e.g., unresolved turbulence, magnetic fields, and cosmic rays) can store additional energy that dissipates over time-scales that are typically longer than the gas cooling time-scale. In order to mimic the effect of non-thermal pressure terms, we turn off the cooling for gas around star particles for a dissipation time-scale $t_{\rm diss}=20\Myr$ \citep[see, e.g.,][]{Gerritsen1997,Thacker_and_Couchman2000,Thacker_and_Couchman2001,Stinson2006,Teyssier2013}, of the order of the life-time of massive ($>10\Msun$) stars. In \simname, we use the same dissipation time-scale for both PopII and PopIII star particles.

\subsection{Lyman-Werner radiation}
The total (unattenuated) Lyman-Werner (LW) flux $J_{\,\mathrm{LW},i}$ at the centre of the $i$-th halo is computed accounting for a uniform LW background flux \citep[$J_{\,\rm LW}^{\,\rm bk}$; following][]{Incatasciato2023} and the LW flux received from star forming regions around the target halo:
\begin{equation}
    J_{\,\mathrm{LW},i} 
    \,=\, J_{\,\rm LW}^{\,\rm bk} \,+\,
    \sum_{n}^{\rm stars}\,\mathcal{A}_{\star, n}\left(\frac{m_{\star,n}}{10^3\Msun}\right)\left(\frac{\left\lvert\bm{r}_i-\bm{r}_{\star,n}\right\rvert}{1\kpc}\right)^{-2}
    ~.\label{eqn:J_LW_0}
\end{equation}
\noindent Here $m_{\star,n}$ and $\bm{r}_{\star,n}$ are the mass and position vector of the $n$-th star particle, respectively. The factor $\mathcal{A}_{\star,n}$ includes the dimming of the LW flux with increasing stellar age $t_\star$ \citep[][]{Schaerer2003,Lupi2021}. For the $n$-th star particle, we have:
\begin{equation}
    \mathcal{A}_{\star,n} = \frac{f_{\star,n}\,\exp\left[\,-t_{\star,n}/\left(300\Myr\right)\,\right]}{\left[\,1+t_{\star,n}/\left(4\Myr\right)\,\right]^{3/2}}
    ~,
\end{equation}
\noindent where $f_{\star,n}=15$ for PopIII stars, and $f_{\star,n}=3$ otherwise \citep[e.g.,][]{Greif_and_Bromm2006,Agarwal2014}.

We correct the LW fluxes of Equation~\eqref{eqn:J_LW_0} for attenuation and shielding, by estimating the molecular gas fraction and column density of haloes along the line-of-sight of every source.

\subsection{Mass-inflow rates}
We estimate the instantaneous mass-flow rate $\dot{M}_i$ onto the central region (where we expect DCBHs to form) of the $i$-th halo as follows:
\begin{equation}
    \dot{M}_{\mathrm{in},\,i} \,=\, -\sum_{n}\,\frac{m_{n}\,v_{r,n}}{r_n}
    ~,\label{eqn:Mdot}
\end{equation}
\noindent where $m_n$, $v_{r,n}$ and $r_n = \left\lvert\bm{r}_i-\bm{r}_{n}\right\rvert$ are the mass, radial component of velocity, and distance, respectively, of the gas in the $n$-th cell, with respect to the centre of the $i$-th halo.

\subsection{Black hole formation}\label{sec:DCBH_model_methods}
At each time-step, for all haloes identified in a given simulation, we also evaluate the physical conditions around the density peak, within a spherical accretion radius $r_{\rm accr} = 4\,\Delta x_{\rm min}$, where $\Delta x_{\rm min}$ is the size of a cell at the highest refinement level in the simulation.

A given halo is considered a candidate DCBH-formation site if all of the following criteria are met:
\begin{itemize}
    \item
    The average density, $\bar{n}$, around the halo's peak exceeds the threshold density for star-formation, $n_{\rm SF}$: $\bar{n}>n_{\rm SF}$.\\
    
    \item
    The average gas-phase metallicity, $\bar{Z}$, around the halo's peak below a defined threshold, $Z^{\,\rm crit}$: $\bar{Z}<Z^{\,\rm crit}$.\\
    
    \item
    The effective Lyman-Werner flux, $J_{\,\rm LW}$, at the halo's peak must exceed a critical value, $J_{\,\rm LW}^{\,\rm crit}$: $J_{\,\rm LW}>J_{\,\rm LW}^{\,\rm crit}$.\\

    \item
    The mass-flow rate, $\dot{M}_{\rm in}$, toward the halo's density peak must exceed a critical value, $\dot{M}^{\,\rm crit}_{\rm in}$: $\dot{M}_{\rm in}>\dot{M}^{\,\rm crit}_{\rm in}$.\\

    \item
    The halo must not have formed stars: $M_{\rm star} = 0$.\\

\end{itemize}

We do not consider any DCBH formation criterion that evaluates the spin of the dark matter halo, because the halo spin is not clearly coupled to the angular momentum of baryons on smaller scales \citep[e.g.,][]{Dubois2012b,Bonoli2014}. Furthermore, a low-spin criterion is generally less restrictive criterion compared to requiring a minimum LW flux \citep[][]{Bhowmick2022a}.

In Table~\ref{tab:runs}, we summarize the criteria used for DCBH formation in the sub-set of \simname simulations that we use in this work, where we assume different values for the relevant critical thresholds mentioned above. \NEW{Specifically, simulation M000 assumes locally adjusted minimum thresholds for both $J_{\rm LW}$ and $\dot{M}_{\rm in}$ and a fixed maximum metallicity threshold. Instead, simulations M002 and M004 use fixed thresholds for all criteria. We use a fixed critical $J_{\rm LW}=100~J_{21}$ for both simulations, to compare to previous works on DCBH formation \citep[e.g.,][]{Habouzit2016,Yue2017,Bhowmick2024c,OBrennan2025}, and explore the theoretical range of critical inflow rates using two fixed thresholds at $0.1\Msun\,\mathrm{yr}^{-1}$ (in M002) and $1\Msun\,\mathrm{yr}^{-1}$ (in M004). Additionally, in simulation M004, we use a lower maximum metallicity of $10^{-6}~Z_\odot$ for DCBH formation. These three simulations were specifically chosen to report results from a reasonable variety of scenarios for DCBH formation. Different combinations of critical thresholds and criteria are explored in the full suite of \simname simulations, with higher resolution and larger simulated volumes.}

\subsubsection{Local critical Lyman-Werner flux}
Assuming fully neutral (i.e., mean molecular weight $\mu=1.22$) monoatomic medium (adiabatic index $\gamma=5/3$) in a virialized halo, we can give an estimate for the local, critical LW flux $J_{\,\mathrm{LW},\,\rm loc}^{\,\rm crit}$ above which the LW flux prevents H$_2$ from efficiently forming in a target halo \citep[see, e.g.,][and references therein]{Schauer2019,Kulkarni2021,Lupi2021}:
\begin{equation}
    \frac{J_{\,\mathrm{LW},\,\rm loc}^{\,\rm crit}}{100\,J_{21}} \;=\; \epsilon_{\rm H_2} \, \left(\frac{1+z}{10}\right)^{2}\,\left(\frac{n}{\mathrm{cm}^{-3}}\right)^2\,\left(\frac{M}{10^{8}\,\mathrm{M}_\odot}\right)^{4}
    ~,\label{eqn:J_LW_crit}
\end{equation}
\noindent where $n$ and $M$ are the average hydrogen density and total mass (dark matter + baryons) of the target halo at redshift $z$, respectively. Dense-enough haloes in our simulations have masses ranging from approximately $10^6$ to $10^{10}\Msun$, implying a vast range of critical LW fluxes. The quantity $\epsilon_{\rm H_2}=t_{\rm cool, \rm H_2}^{\,\rm min}/t_{\rm H}$ is the minimum H$_2$ cooling time-scale $t_{\rm cool, \rm H_2}^{\,\rm min}$ allowed for DCBH formation, in units of the Hubble time-scale $t_{\rm H}$. In \simname, we assume $\epsilon_{\rm H_2}=1$. In order to be considered a candidate DCBH-formation site we ask the target halo to satisfy $J_{\,\mathrm{LW}}^{\,\rm crit}>J_{\,\mathrm{LW},\,\rm loc}^{\,\rm crit}$ or $100~J_{21}$, depending on the simulation (see Table~\ref{tab:runs}).

\subsubsection{Dynamical heating}
Dynamical, compressional heating can counteract cooling and prevent H$_2$ formation for halo mass-growth rates above a local threshold value $\dot{M}^{\,\rm crit}_{\,\rm in,\rm loc}$ \citep[see, e.g.,][and references therein]{Yoshida2003,Wise2019,Lupi2021}. Furthermore, sustained inflow rates onto the central regions of DCBH-forming haloes are required to fuel the accretion of the intermediate protostar phase and overcome the action of radiative feedback \citep[e.g.,][]{Hosokawa2012,Hosokawa2013,Schleicher2013,Chon2018}. Following, e.g., \citet[][]{Lupi2021}, for a target halo at redshift $z$ we have:
\begin{equation}
   \frac{\dot{M}^{\,\rm crit}_{\,\rm in, \rm loc}}{\mathrm{M}_\odot/\mathrm{yr}}
   \,=\,\left(\frac{1+z}{10}\right)^{3/2}\left(\frac{M}{10^8\,\mathrm{M}_\odot}\right)^{-1/2}\left(\frac{\epsilon_{\rm H_2}\,J_{\,\mathrm{LW}}}{J_{\,\mathrm{LW},\,\rm loc}^{\,\rm crit}}\right)^{-1}
   ~,\label{eqn:Mdot_crit}
\end{equation}
\noindent where $n$, $M$, $J_{\,\mathrm{LW}}$ are the average hydrogen density, total mass (dark matter + baryons), and received LW flux of the target halo at redshift $z$. In order to be considered a candidate DCBH-formation site, the target halo must satisfy either $\dot{M}_{\rm in}>\dot{M}^{\,\rm crit}_{\,\rm in,\rm loc}$, $0.1~\Msun\,\mathrm{yr}^{-1}$, or $1~\Msun\,\mathrm{yr}^{-1}$, depending on the simulation (see Table~\ref{tab:runs}).

\subsubsection{Seed mass}
Unlike most state-of-the-art simulations \citep[e.g., Illustris, EAGLE; but see][]{Habouzit2017}, the initial mass of DCBHs in \simname is determined by local properties of their host haloes, rather than being set to a fixed value.

All clumps that satisfy the above criteria are flagged as DCBH-formation sites. \NEW{To emulate DCBH formation over the timescales during which the formation site is expected to experience intense UV radiation, strong inflows, and accretion during the intermediate supermassive star phase, at} each time-step $\Delta t$, we assign the DCBH-formation site a Poisson probability $\mathcal{P}_{\bh}$ of forming a DCBH sink particle that time-step \citep[e.g.,][]{Bellovary2011,Dunn2018}:
\begin{equation}
    \mathcal{P}_{\bh} = 1 - \exp\left(-\Delta t / t_{\mathrm{ff}}\right)
    ~,
\end{equation}
\noindent where $t_{\mathrm{ff}} = \sqrt{3\pi/32\,G\,\bar{\rho}\;}$ is the local free-fall time in the halo. Here $\bar{\rho}$ is the average gas mass density in the cells around the DCBH-formation site, computed within $r_{\rm accr}$. For densities above the star-formation threshold of \simname, we have that $t_{\rm ff}\lesssim 70\Myr$. We then draw a uniform random number $x\in \left[0,1\right]$ and, if $x<\mathcal{P}_{\bh}$, we form a DCBH sink particle with initial mass $M_{\bh}^{\,\rm seed}$ given by:
\begin{align}
    M_{\bh}^{\,\rm seed} \,
    &=\,\bigg{\{}2.74\,+\,2.21\,\log_{10}\left(\frac{\langle\,\dot{M}\,\rangle_{\bh}}{\mathrm{M}_\odot\,\mathrm{yr}^{-1}}\right)\,
    \nonumber\\
    &+\,0.64\,\left[\log_{10}\left(\frac{\langle\,\dot{M}\,\rangle_{\bh}}{\mathrm{M}_\odot\,\mathrm{yr}^{-1}}\right)\right]^2\bigg{\}}\times 10^5\Msun
    ~,\label{eqn:BH_ini}
\end{align}

\noindent where $\dot{M}$ is the mass-inflow rate computed as in Equation~\eqref{eqn:Mdot}, and $\left\langle\,\cdot\,\right\rangle_{\bh}$ represents the volume-weighted average computed using a cloud-in-cell (CIC) interpolation over the 8 nearest father cells around the DCBH-formation site. Equation~\eqref{eqn:BH_ini} above is derived as the average fit to the results of \citet[][]{Umeda2016,Woods2017,Haemmerle2018} for the final mass of accreting (non-rotating) supermassive stars in pristine cores \citep[see also][for a review]{Woods2019} with a second-order polynomial.

A swarm of \textit{cloud} particles is spawned around every sink particle, spaced by $\Delta x/2$ and within a sphere of radius $r_{\rm cloud} = 4 \Delta x$, where $\Delta x$ is the local cell size. \NEW{The role of cloud particles is purely numerical. They serve to constantly probe the average physical properties of gas around BHs, to then allow us to calculate the relevant quantities used for, e.g., our BH growth schemes (see, e.g., in Equation~\ref{eqn:Bondi} below).}

\begin{figure*}
    \centering
    \includegraphics[width=.29\hsize]{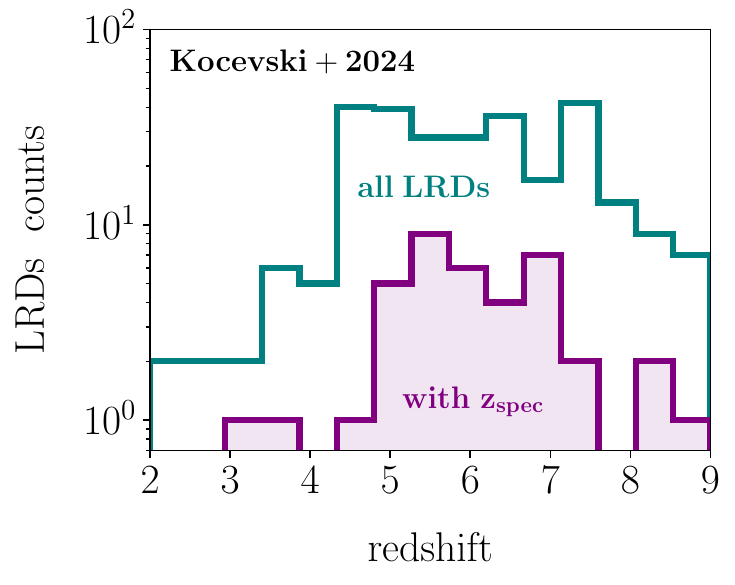}
    \includegraphics[width=.35\hsize]{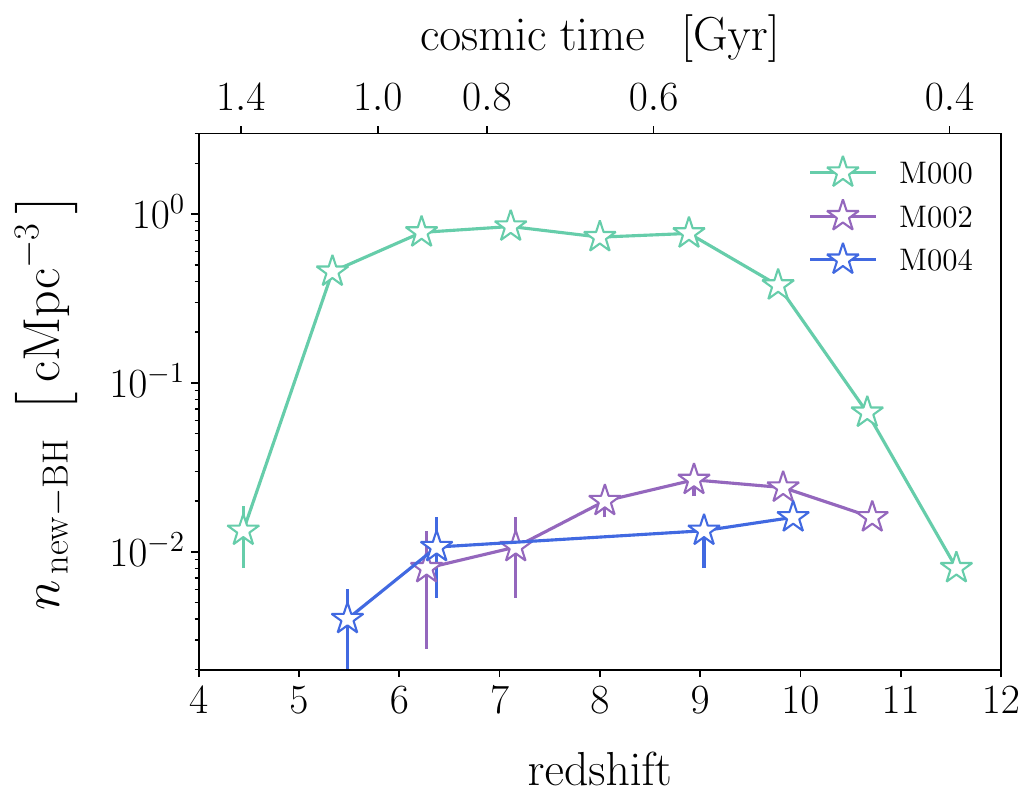}
    \includegraphics[width=.35\hsize]{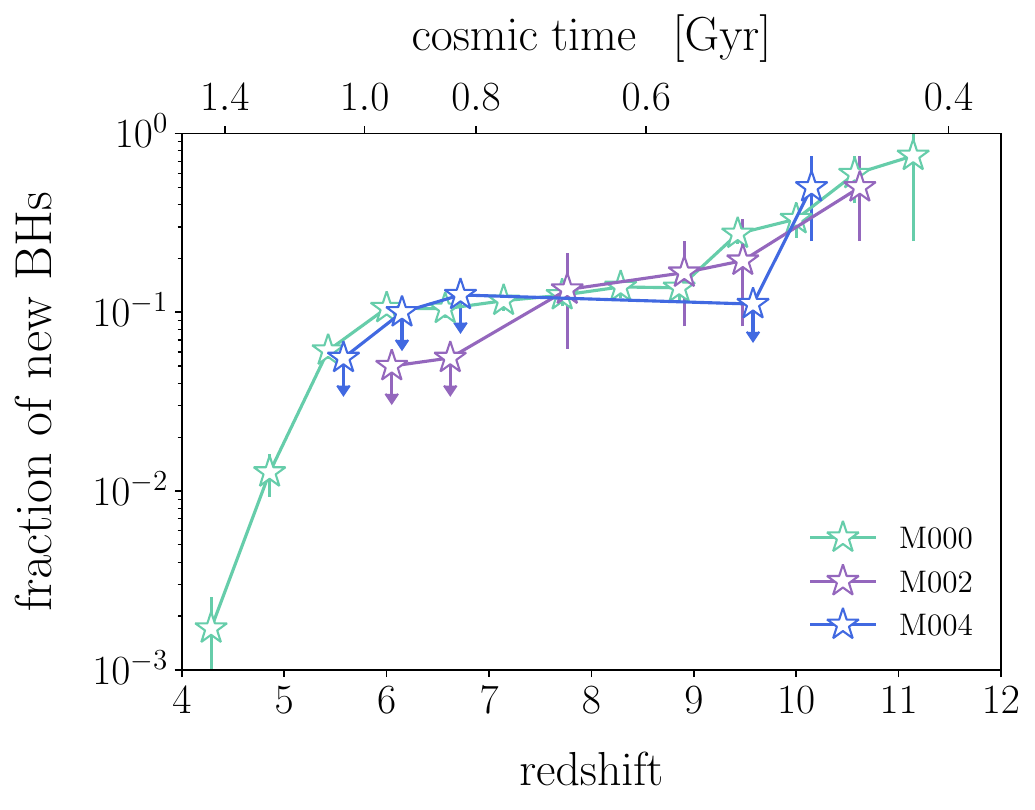}
    \caption{\textit{Left panel}: redshift distribution of all LRDs reported by \citet[][]{Kocevski2024} \citep[see also Figure 1 of][]{Inayoshi2025}. The filled, purple histogram shows the sub-set of LRDs with spectroscopically confirmed redshift ($z_{\rm spec}$). \textit{Middle panel}: abundance of newborn DCBHs ($n_{\rm new-BH}$; BH age $<100\Myr$) as function of redshift, in different simulations. \textit{Right panel}: fraction of newborn DCBHs compared to the total number of DCBHs present in the simulated volumes at the same redshift, as a function of redshift. Data points are the median values with bootstrapped (16-th to 84-th percentiles) errorbars. The formation of DCBHs appear to be significantly less efficient at $z\lesssim 6$, as the fraction of newly formed DCBHs declines steeply. This behaviour is akin to that observed in the emergence of LRDs \citep[][]{Kocevski2024,Inayoshi2025}. Different simulations yield different overall abundances but consistent fractions of newborn DCBHs. Simulations with more stringent, fixed thresholds for DCBH formation (i.e. M002 and M004) possibly display a declining population of newborn DCBHs already at $z>7$. The volume of our simulations limits the measurable \NEW{instantaneous} abundances to $>8\times 10^{-3}~\mathrm{cMpc}^{-3}$, making it more difficult to understand the behaviour of the simulations when no DCBHs from.}
    \label{fig:BH_abundance}
\end{figure*}

\begin{figure*}
    \centering
    \includegraphics[width=\hsize]{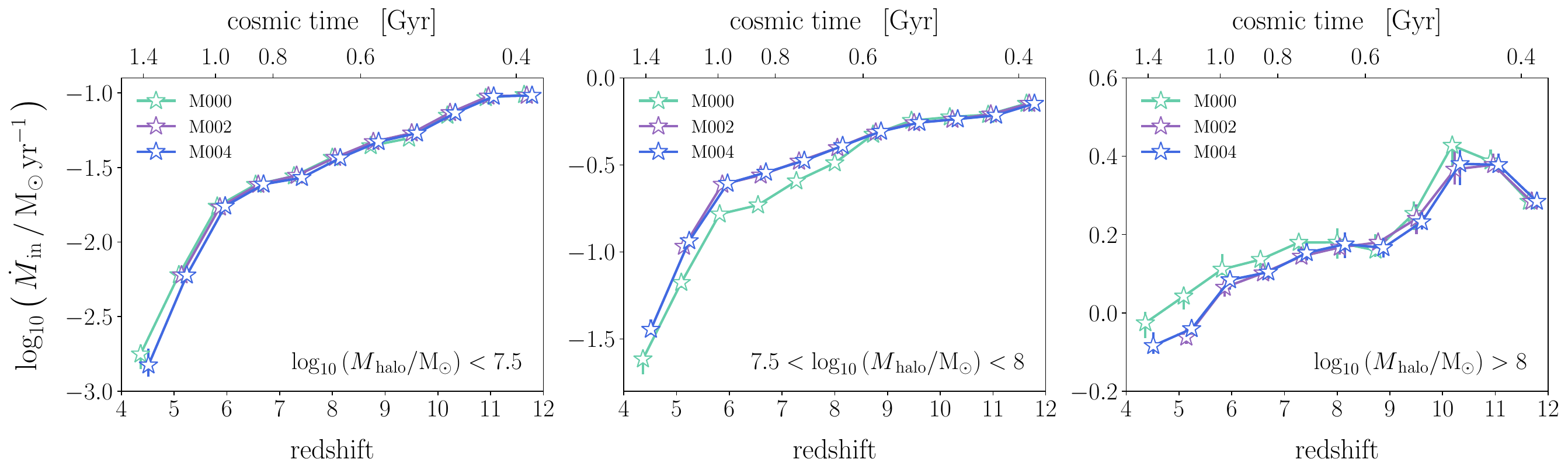}
    \caption{Mass-inflow rates ($\dot{M}_{\rm in}$) in \textit{all} haloes \NEW{(whether they host a DCBH or not)} in our simulations, as a function of redshift, in different simulations. We show these results for three different halo mass bins, increasing from left to right. Data points are the median values with bootstrapped (16-th to 84-th percentiles) errorbars. In general, the median mass-inflow rate of haloes of a given mass decreases at lower redshifts, displaying a steep decline at $z\lesssim 6$ (especially at halo masses $M_{\rm halo}\lesssim 10^8\Msun$) that correlates with the decline in the fraction of newborn DCBHs. In this context, different simulations produce very similar results. In general, the less efficient formation of DCBH at low redshifts is due to haloes being progressively more metal-enriched, with less dense gas, more stars, and inefficient inflows \NEW{(see Appendix~\ref{app:haloes} for details on the evolution of physical properties of haloes, other than inflow rates)}.}
    \label{fig:inflow}
\end{figure*}

\begin{figure}
    \centering
    \includegraphics[width=\hsize]{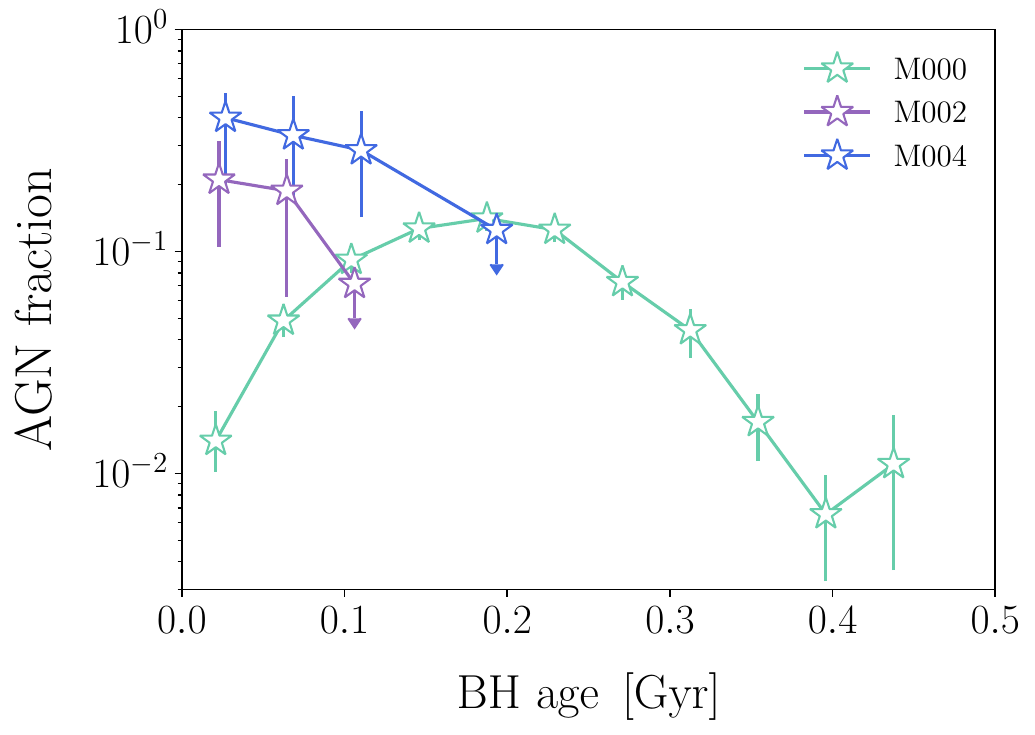}
    \caption{Fraction of all DCBHs of a given age (i.e. time since their formation; $\mathrm{BH\,age}$) in the simulated volume that have a bolometric luminosity $L_{\rm bol}>10^{43}~\mathrm{erg}\,\mathrm{s}^{-1}$ (i.e. classified as AGN), in different simulations. Data points are the median values with bootstrapped (16-th to 84-th percentiles) errorbars. In general, DCBHs do exhibit an AGN phase in the $\lesssim 200\Myr$ after their formation. The overall AGN fraction and the characteristic time-scales associated with the AGN phase do differ in different simulations.}
    \label{fig:AGN_fraction}
\end{figure}

\begin{figure*}
    \centering
    \includegraphics[width=.45\hsize]{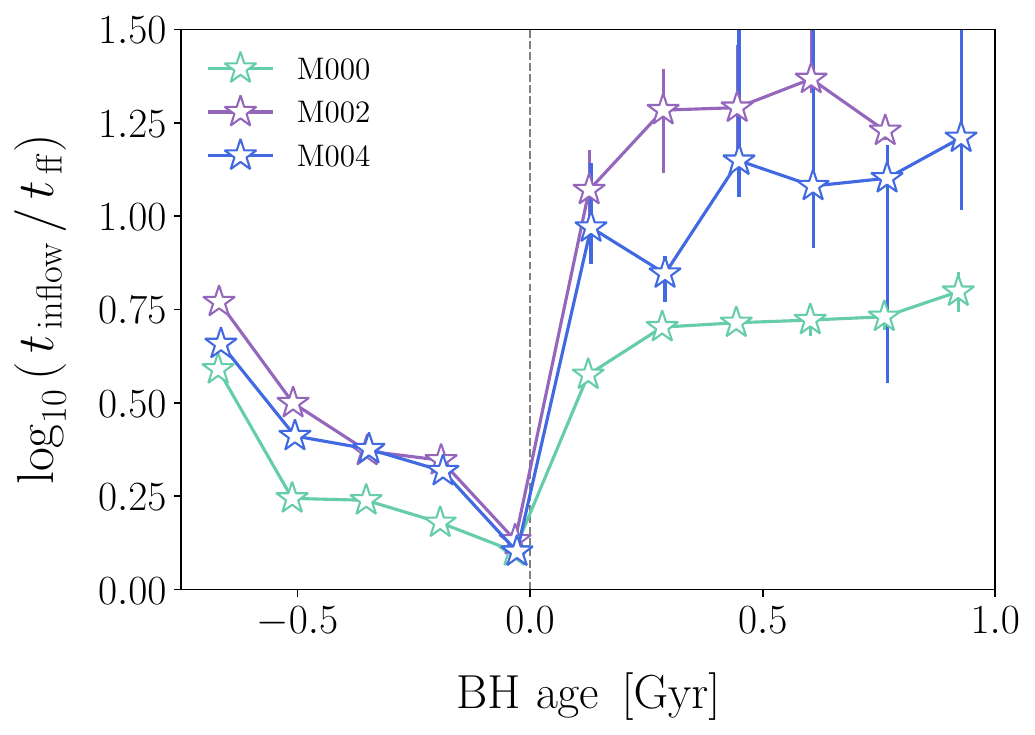}
    \includegraphics[width=.45\hsize]{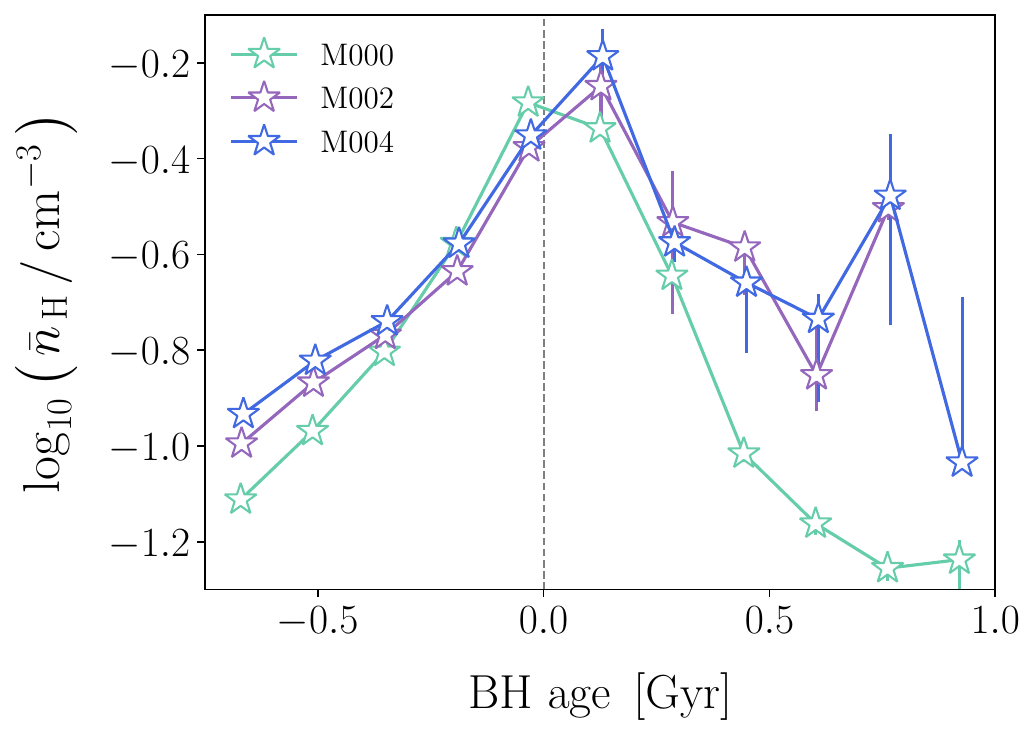}
    \includegraphics[width=.45\hsize]{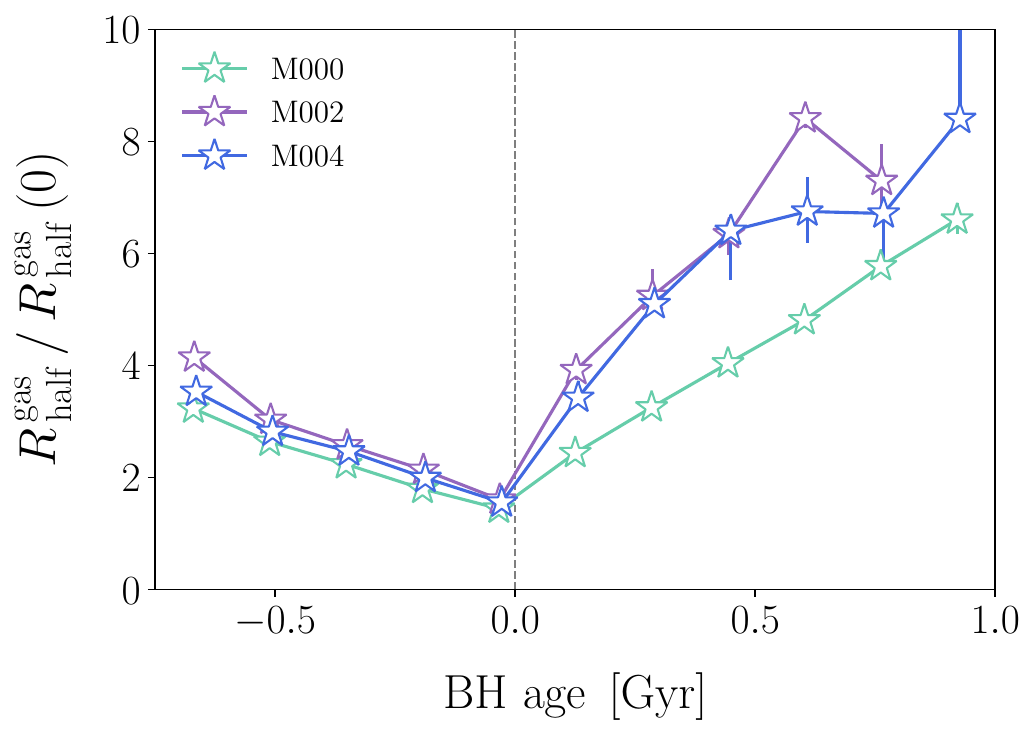}
    \includegraphics[width=.45\hsize]{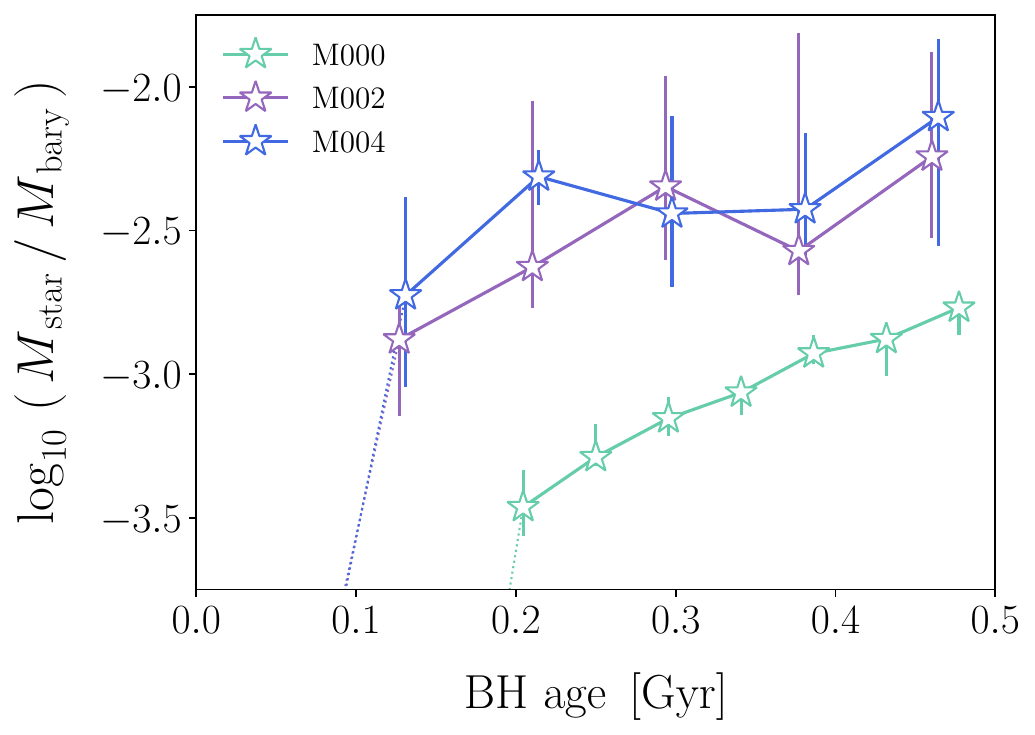}
    \caption{Evolution of the properties of haloes hosting DCBHs, prior to and after their formation, as function of the BH age (time since the DCBH have formed), in different simulations. Data points are the median values with bootstrapped (16-th to 84-th percentiles) errorbars. In the upper left panel, we show the evolution of the ratio between the inflow time-scale ($t_{\,\rm inflow}\equiv M_{\rm gas}/\dot{M}_{\rm in}$; i.e. the ratio between the total gas mass available and the mass-inflow rate in the halo) and the local free-fall time-scale ($t_{\rm ff}$). In the uppper right panel, we show the evolution of the average hydrogen number density ($\bar{n}_{\,\rm H}$) in the halo. In the bottom left panel, we show the evolution of the gas half-mass radius ($R_{\,\rm half}^{\,\rm gas}$), where the trend for each halo was normalized by $R_{\,\rm half}^{\,\rm gas}$ at the time of DCBH formation ($R_{\,\rm half}^{\,\rm gas}\left(0\right)$; i.e. when $\mathrm{BH\,age}=0$). For $\mathrm{BH\,age}=0$, the data points do not meet at $R_{\,\rm half}^{\,\rm gas}/R_{\,\rm half}^{\,\rm gas}(0)=1$, but rather at a larger value, due to the choice of binning in BH age. In the bottom right panel, we show the evolution of the stellar mass fraction ($M_{\rm star}/M_{\rm bary}$; i.e. stellar mass over total baryonic mass). The general picture is that efficient inflows are associated with a gas compaction event and the formation of a DCBH. Subsequently, on average, inflows get less efficient, gas becomes more diffuse, and the first PopIII stars can form.}
    \label{fig:gas_evolution}
\end{figure*}

\begin{figure}
    \centering
    \includegraphics[width=\hsize]{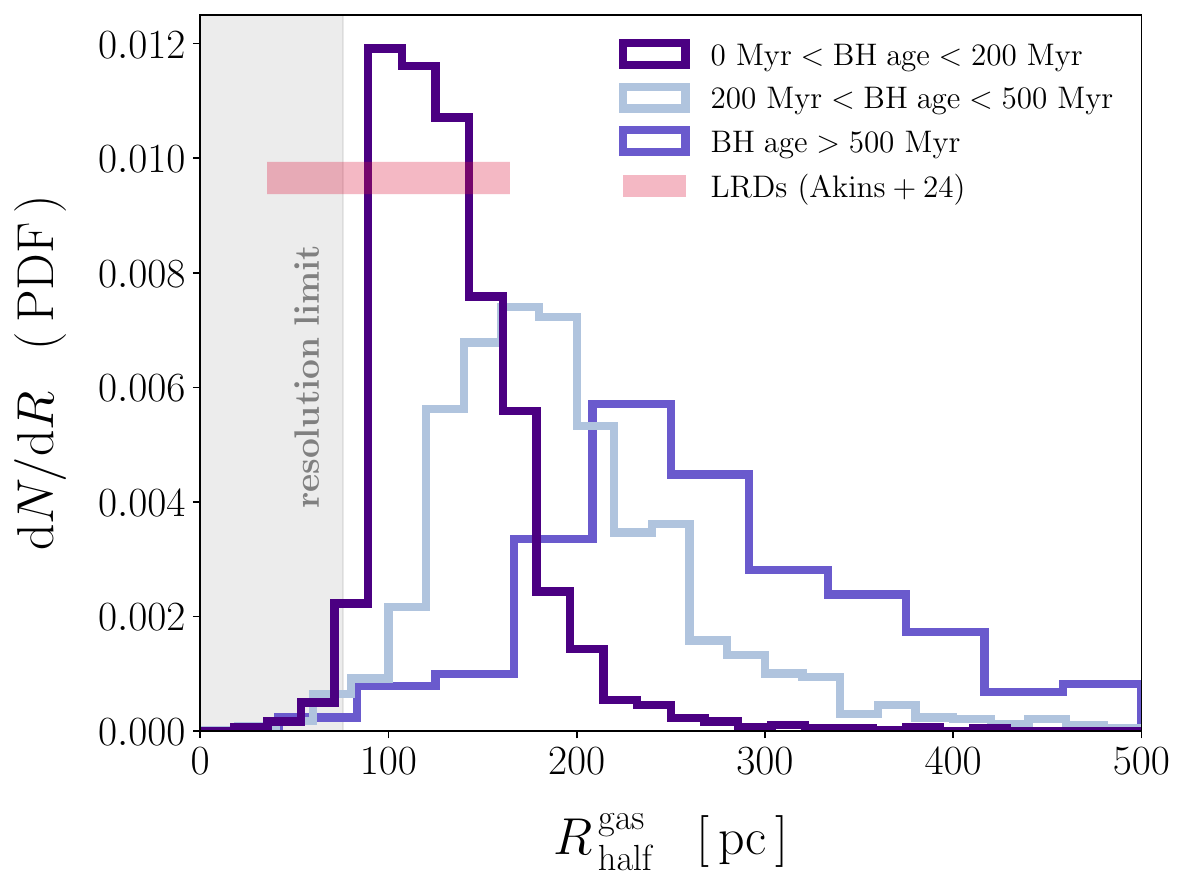}
    \caption{Estimated probability density function (PDF) for the half-mass radius ($R_{\,\rm half}^{\,\rm gas}$) of gas in haloes that host a DCBH, for different BH ages, in all simulations. \NEW{The red shaded area represents the 16th-to-84th percentiles range of estimated effective radii for the LRDs by \citet[][F444W NIRCam filter]{Akins2024}.} In general, newly-formed DCBHs ($\mathrm{BH\,age}\lesssim 200\Myr$) are embedded in dense gas reservoirs with half-mass radii $\lesssim 100-200\pc$.}
    \label{fig:Rhalf}
\end{figure}

\begin{figure}
    \centering
    \includegraphics[width=\hsize]{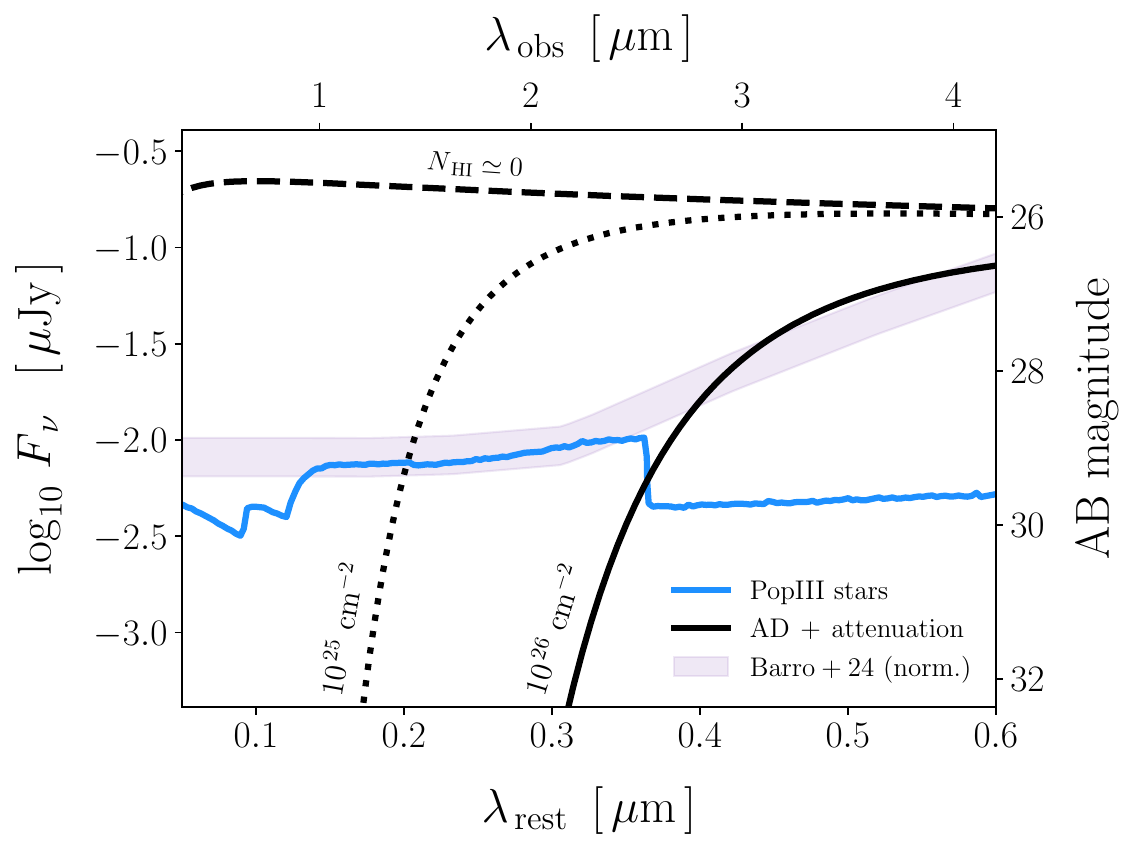}
    \caption{Illustration of the spectral energy distribution (SED) expected from the multi-temperature alpha-accretion disc \citep[black lines;][]{Shakura_and_Sunyaev1973} around a $3\times 10^7\Msun$ BH accreting at the Eddington limit (bolometric luminosity $L_{\rm bol}\simeq 4\times 10^{45}~\mathrm{erg/s}$) and a population of $3\times 10^6\Msun$ zero-age PopIII stars \citep[blue line; Salpeter initial-mass function; ][]{Schaerer2002} in a DCBH ‘nursery' at $z=6$. Attenuation is computed considering the contribution from Rayleigh scattering and dust extinction for different neutral hydrogen column densities ($N_{\rm HI}=0,\,10^{25},\,\mathrm{and}\,10^{26}~\mathrm{cm}^{-2}$; for the dashed, dotted, and solid black line, respectively) and a gas-phase metallicity $Z=10^{-6}~Z_\odot$ (see main text for details). For comparison, we show the average SED from the sample of LRDs from \citet[][]{Barro2024} ($z=5.5,\,\text{and}\,7.5$), re-normalized by the value of our total (stars + AGN with $N_{\rm HI}=10^{26}~\mathrm{cm}^{-2}$) SED at $0.4~\mu\mathrm{m}$.}
    \label{fig:LRD_SED}
\end{figure}
\subsection{Black hole growth}
DCBHs in the \simname simulations grow due to both black hole mergers and gas accretion, that will affect the evolution of their mass and spin.

The black hole mass growth rate due to gas accretion is estimated starting from the Bondi-Hoyle-Lyttleton \citep[][]{Hoyle_and_Lyttleton1939,Bondi_and_Hoyle1944,Bondi1952} solution:
\begin{equation}
    \dot{M}_{\rm B} \,=\,\frac{4\,\pi\,G^2\,M_{\bh}^2\,\left\langle\rho\right\rangle_{\bh}/\alpha_{\rm B}}{\left[\;\left\lvert \bm{v}_{\bh}-\left\langle\bm{v}\right\rangle_{\bh}\right\rvert^2 + \left\langle c^2_{\rm s}\right\rangle_{\bh}\;\right]^{3/2}}
    ~,\label{eqn:Bondi}
\end{equation}
\noindent where $M_{\bh}$, and $\bm{v}_{\bh}$ are the black hole mass and velocity, respectively, whereas $\bm{v}$, $c_{\rm s}$, and $\rho$ are the velocity, sound speed, and mass density of gas in the cells surrounding the sink particle, respectively. With $\left\langle \cdot\right\rangle_{\bh}$ we represent the volume-weighted average with CIC interpolation over the 8 nearest father cells around the sink particle, within $r_{\rm cloud}$. The correction factor $\alpha_{\rm B}$ is introduced to extrapolate the average gas quantities to their value far from the black hole \citep[see][]{Krumholz2004,Dubois2010}. 


Furthermore, we limit the accretion rate $\dot{M}_{\rm B}$ to the Eddington critical accretion rate $\dot{M}_{\rm Edd} = 4\pi\,G\,M_{\bh}\,m_{\rm H}/\left(\effrad\,\sigma_{\rm T}\,c\right)$, where $\sigma_{\rm T}$ is the Thomson scattering cross-section, and $\effrad$ is the radiative efficiency of the accretion process. To summarize, the mass-flow rate towards the black hole is $\dot{M}_{\rm accr}=\min\{\dot{M}_{\rm B},\dot{M}_{\rm Edd}\}$ and the effective black hole mass-growth rate $\dot{M}_{\bh}=\left(1-\effrad\right)\,\dot{M}_{\rm accr}$. Over a time-step $\Delta t$, the black hole mass grows of $\dot{M}_{\bh}\,\Delta t$ and a mass $\dot{M}_{\rm accr}\,\Delta t$ is removed from the gas in the cells within $r_{\rm accr}$ from the sink particle. The amount of mass removed from each cell follows the same volume-weighted scheme as the one used to compute average physical quantities in the vicinity of the sink particle.

From the disc mass-flow rate, $\dot{M}_{\rm accr}$, we can estimate the bolometric luminosity produced in the accretion process as follows:
\begin{equation}
    L_{\rm bol} \,=\, \effrad\,\dot{M}_{\rm accr}\,c^2
    ~.\label{eqn:Lbol_accr}
\end{equation}
\noindent In this work, we assume an effective radiative efficiency $\effrad=0.1$ when calculating bolometric luminosities \NEW{and accretion rates.}.

\subsection{Black hole feedback}
We include two modes of BH feedback that are effective in different regimes, characterised by their Eddington ratio $\lambda_{\rm Edd}\equiv \dot{M}_{\rm accr}/\dot{M}_{\rm Edd}$: a ‘radio mode' for low $\lambda_{\rm Edd}<0.01$ and a ‘quasar mode' for higher $\lambda_{\rm Edd}$. In the quasar mode, we isotropically inject thermal energy in the medium surrounding the BH, at a rate proportional to the BH bolometric luminosity, $\dot{E}_{\rm AGN}=\epsilon_{\rm q}\,L_{\rm bol}$, with a coupling efficiency $\epsilon_{\rm q}=0.9$ \citep[][]{Booth_and_Schaye2009,Dubois2012a}. In the radio mode, we transfer momentum to the gas around the BH, as a bipolar outflow with jet velocity of $v_{\rm jet}=10^4~\mathrm{km/s}$ \citep[with a mass-loading factor $\eta_{\rm kin}=100$ of the jet on unresolved scales; see, e.g.,][]{Omma2004,Cattaneo_and_Teyssier2007,Dubois2009,Dubois2010}. Kinetic energy and momentum are spread over a cone of aperture $\theta_{\rm cone}=90~\mathrm{deg}$, aligned with the average angular momentum of the gas around the BH.

\section{Declining population of newborn DCBHs}\label{sec:decline}
In the next sections, we analyse the results from the three simulations introduced above (M000, M002, and M004; see a summary in Table~\ref{tab:runs}). We primarily focus on the evolution of the population of newly-formed DCBHs and the properties of their host haloes around the time of DCBH formation.

DCBHs in \simname form with initial masses in the range $10^4 - 10^6\Msun$. By redshift $z=7$, we form about 300 DCBHs (with a corresponding abundance of $2.44~\mathrm{cMpc}^{-3}$) in simulation M000, whereas in M002 and M004 we only form 8 ($0.064~\mathrm{cMpc}^{-3}$) and 3 ($0.024~\mathrm{cMpc}^{-3}$), respectively. This differences owe to different formation criteria employed in the different simulations. Specifically, simulation M000 \NEW{allows for enhanced DCBH formation in haloes with masses $\lesssim 10^8\Msun$, where the critical LW flux (Equation~\ref{eqn:J_LW_crit}) is significantly smaller than in higher mass haloes. Depending on the total received LW flux, the critical inflow rate could also reduce in low mass haloes (see Equation~\ref{eqn:Mdot_crit})}.

In Figure~\ref{fig:BH_abundance}, we show the abundance of newborn DCBHs (i.e. with ages $\mathrm{age_{\bh}}<100\Myr$) as function of redshift, in different simulations (middle panel), and the fraction of newborn DCBHs compared to the total number of DCBHs that are present in the simulated volume at a given redshift, in different simulations (right panel). In \NEW{simulation M000}, both the abundance and fraction of newly formed DCBHs \NEW{exhibits a clear, steep decline} at $z\lesssim 6$, indicating that DCBH formation becomes significantly less efficient at lower redshifts. Interestingly, this behaviour is akin to that observed in the emergence of LRDs \citep[left panel;][]{Kocevski2024}. \NEW{In simulations M002 and M004, no new BHs are formed after $z\simeq 6$ and $z\simeq 5$, respectively, albeit our low statistics introduces a large uncertainty when constraining their abundance at $z\lesssim 8$, especially in M004.} The different simulations yield different overall abundances but consistent fractions of newborn DCBHs, especially at $z>7$. Simulations with more stringent, fixed thresholds for DCBH formation (i.e. M002 and M004; see Table~\ref{tab:runs}) could have a declining population of newborn DCBHs already \NEW{at $z\gtrsim 9$}. However, the volume of our simulations does not allow us to measure \NEW{instantaneous} abundances below $8\times 10^{-3}~\mathrm{cMpc}^{-3}$, making it more difficult to understand the behaviour of these simulations when no DCBHs from. \NEW{Here, abundances are computed as the number density of newborn BHs averaged over the simulation outputs within the given redshift bin. Consequently, the reported values can lie below the limiting instantaneous measurable abundance in the simulated box. In simulation M004, there is about one newly formed BHs in every considered redshift bin within $z=5-11$. Therefore, we can only provide an interpolation between redshift bins with at least one new BH forming (see absent data points at $z\simeq 7-8$).} Furthermore, DCBHs could form at $z>12$, however, our sample is likely incomplete at higher redshifts, due to the relatively small volume of the simulations presented here. Larger volumes will be simulated to fully probe the cosmic number density of DCBHs (see Cenci et al., in preparation).

Any change in the population of newborn DCBHs correlates, by construction, with changes in the physical state of gas in the target haloes for their formation. In general, low-redshift haloes are more metal enriched, have more stars, and their central gas density is lower than in high-redshift haloes \NEW{(see Appendix~\ref{app:haloes})}. As a result, we expect the formation of DCBHs to become highly inefficient at low redshift. Furthermore, mass-inflow rates are significantly reduced going to lower redshifts. In Figure~\ref{fig:inflow}, we show the median mass-inflow rates in all haloes, as a function of redshift, in different simulations, and in three different halo mass bins. On average, the mass-inflow rate of all haloes decreases with redshift, with a steeper decline at $z\lesssim 6$, that is especially evident for halo masses $M_{\rm halo}\lesssim 10^8\Msun$. This feature in the evolution of inflows correlates with the decline in the fraction of newborn DCBHs that we see in Figure~\ref{fig:BH_abundance}.

Our results suggests that the decline in the fraction of new DCBHs (or, equivalently, in the DCBH-formation rate) is associated with the decreasing efficiency of inflows within haloes. This behaviour is likely associated with rapidly decreasing gas fractions and reduced cold accretion in our simulations, at $z\lesssim 6$ \citep[see, e.g.,][]{Sargent2014,Miettinen2017,Liu2019}. Inflow rates within low-mass haloes can correlate with the accretion rates from larger scales onto haloes, and the reduced inflows would allow for an effective increase in the gas-phase metallicity of haloes \citep[e.g.,][]{Dave2012,Lilly2013,Bassini2024}, hence hindering the formation of DCBHs. Furthermore, increasing metallicities enhance the effect of feedback in driving outflows and therefore reducing the effective inflow rates within haloes \citep[e.g.,][]{Muratov2015,Bassini2023}. The relatively low inflow rates within low-mass haloes at $z\lesssim 6$ can still be above the assumed critical thresholds, and all DCBH formation criteria may be important in determining the evolution of the abundance of newly-formed DCBHs. \NEW{In Appendix~\ref{app:haloes}, we show how central gas densities decrease and, on average, gas-phase halo metallicities increase, with redshift, effectively reducing DCBH formation.} Understanding the causal connection between the decline in DCBH formation and the physical state of galaxies is beyond the scope of this work. A comprehensive analysis of the halo and galaxy properties in our simulations is deferred to a companion paper (Cenci et al., in preparation).

\section{Before and after DCBH formation}\label{sec:evolution}

In Figure~\ref{fig:AGN_fraction}, we show the fraction of DCBHs in our simulations that is classified as AGN (i.e. with bolometric luminosities $L_{\rm bol}>10^{43}~\mathrm{erg}\,\mathrm{s}^{-1}$) as a function of the time since the DCBH formation (BH age). All DCBHs contributing to this AGN fraction are accreting close to their Eddington limit ($<10^{44}~\mathrm{erg}\,\mathrm{s}^{-1}$ for DCBH masses $<10^6\Msun$). In general, a non-negligible fraction of DCBHs is in a luminous AGN phase only in the $\sim 100-200\Myr$ after their formation, depending on the simulation.

Most LRDs have bolometric luminosities $\gtrsim 10^{43}~\mathrm{erg}\,\mathrm{s}^{-1}$ \citep[e.g.,][]{Harikane2023,Akins2024,Greene2024,Kokorev2024,Maiolino2024,Matthee2024a,Ubler2024,Lin2025}. However, only DCBHs during their AGN phase, representing a small fraction of all DCBHs in our simulations, have such high luminosities. Furthermore, we do not model super-Eddington accretion in our simulations, imposing a limit on the maximum luminosity of accreting DCBHs (about $10^{43}~\mathrm{erg}\,\mathrm{s} ^{-1}$ for a BH mass $\sim 10^5\Msun$). In particular, if we only consider luminous newborn DCBHs with $L_{\rm bol}>10^{44}~\mathrm{erg}\,\mathrm{s} ^{-1}$, their abundance would be at least factor $\lesssim 10^{-3}$ lower than the one reported in Figure~\ref{fig:BH_abundance}, in reasonable agreement with the reported abundance of LRDs \citep[e.g.,][]{Kocevski2024}.

In Figure~\ref{fig:gas_evolution}, we show the time evolution of the properties of the host haloes of newborn DCBHs, in a period of time prior to and after DCBH formation. In particular, we show the evolution of their inflow characteristic time-scale, average hydrogen density, half-mass radius, and stellar mass fraction. In general, prior to DCBH formation, haloes experience a clear gas compaction event \citep[typically associated with intense inflows; akin to low-redshift starbursts; e.g.,][]{Dekel_and_Burkert2014,Lapiner2023,Cenci2024a,Cenci2024b,McClymont2025a,McClymont2025b}, characterized by short inflow time-scales (of order of the local free-fall time), increasing gas densities, and decreasing gas half-mass radii. After the formation of the DCBH, inflows rapidly become less efficient (with inflow time-scales a factor $10$ longer than the local free-fall time), density decreases, and the gas reservoir grows in size. Furthermore, the first PopIII stars form in DCBH-host haloes in the $100-200\Myr$ after DCBH formation. 

In Figure~\ref{fig:Rhalf}, we show the typical size of the gas reservoir around DCBHs (in terms of gas half-mass radius, $R_{\rm half}^{\rm gas}$) of different ages. In the $\lesssim 200\Myr$ after DCBH formation, we have gas half-mass radii of about $100-200\pc$, consistently to what is inferred for most LRDs \citep[e.g,][]{Furtak2023,Baggen2024,Casey2024,Labbe2025}. However, we note that these length-scales are close to the spatial resolution (i.e. maximum grid refinement scales) of our simulations and we are thus unable to resolve possibly more compact configurations. After DCBH formation, the gas reservoirs become more extended in size.

\section{Spectrum of DCBH nurseries}\label{sec:SED}
In our simulations, PopIII stars form $\lesssim 200\Myr$ after DCBHs do, when a (small) fraction of the latter are still in their AGN phase. Emission from both newly-formed PopIII stars and the AGN contribute to the intrinsic SED of these systems. To reproduce the characteristic shape of observed spectra of LRDs, we likely need strong dust obscuration, to primarily attenuate the emission from the AGN. However, haloes around our newly-formed DCBHs are practically pristine in terms of average metallicities, by construction (where DCBHs form before their host galaxies), resulting in very low attenuation in the UV. Rayleigh scattering effectively reduces the UV contribution from the AGN only for extremely high HI column densities. Figure~\ref{fig:LRD_SED} shows a simple\footnote{Here we do not include the contribution from either emission lines or scattered light from the AGN, that could boost the overall emission of the systems we considered \citep[e.g.,][]{Leung2024}. Furthermore, the PopIII stars may form following a different initial-mass function, that could also affect the shape and magnitude of the predicted SED \citep[e.g.,][]{Trussler2023}.} SED for a $3\times 10^6\Msun$ zero-age PopIII stellar population \citep[][]{Schaerer2002}, and an optically thick, geometrically thin, standard accretion disc around a $3\times 10^7\Msun$ BH accreting at the Eddington limit \citep[][]{Shakura_and_Sunyaev1973}. We account for effective UV attenuation of the accretion disk emission from Rayleigh scattering with an optical depth \citep[see, e.g.,][]{Lee2013}:
\begin{equation}
    \tau_{\rm scatt} = \left(\frac{1216~\AAA}{\lambda_{\rm rest}}\right)^{4}\,\left(\frac{N_{\rm HI}}{4\times 10^{23}\,\mathrm{cm}^{-2}}\right)
    ~.
\end{equation}
\noindent Furthermore, we consider a dust extinction law \citep[e.g.,][]{Temple2021}:
\begin{equation}
    \tau_{\rm dust} = \frac{A_{\rm V}}{1.086\,R_{\rm V}}\left[R_{\rm V}\,+\,6.2\,+\,\left(\frac{800~\AAA}{\lambda}\right)\right]
    ~,
\end{equation}
\noindent with an extinction ratio $R_{\rm V}=2.7$ \citep[typical value derived for the Small Magellanic Cloud and low-metallicity systems; e.g.,][]{Prevot1984,Calzetti2000}, and $A_{\rm V}=\left(Z/Z_\odot\right)\,\left(N_{\,\rm HI}/1.8\times 10^{21}~\mathrm{cm}^{-2}\right)$ \citep[e.g.,][]{Bohlin1978}, given by the neutral hydrogen column density $N_{\,\rm HI}$ and gas-phase metallicity $Z=10^{-6}~Z_\odot$. Dust extinction only plays a minor role in the overall attenuation, compared to Rayleigh scattering, for $Z\lesssim 10^{-6}~Z_\odot$.

The total optical continuum emission from the majority of DCBH nursery systems would be relatively faint (AB magnitude $>30$) and possibly below the average sensitivity limit of JWST. However, deep JWST surveys \citep[e.g., NGDEEP,GLIMPSE,][]{Bagley2023,Fujimoto2025}, and possibly the addition of strong lensing, should be able to observe them. In our simulations, only $\lesssim 2\percent$ of DCBHs reach masses $\gtrsim 3\times 10^7\Msun$ in the $200\Myr$ after their formation, growing close to their Eddington limit. As a result the abundance of luminous DCBH nurseries could be in agreement with that reported for LRDs (as discussed also in Section~\ref{sec:evolution} concerning the AGN fraction).

We consider different atomic hydrogen column densities, $N_{\rm HI}=0,\,10^{25},\,\text{and}\,10^{26}~\mathrm{cm}^{-2}$. Such high column densities imply strong extinction in the soft X-ray spectrum \citep[e.g.,][]{Ricci2014} and would possibly explain the lack of X-ray detections \citep[e.g.,][]{Ananna2024,Juodzbalis2024,Yue2024,Maiolino2025a}. Precise observations of polarization features around the Lyman-$\alpha$ emission can probe Rayleigh-scattered light off neutral hydrogen gas surrounding these AGNs \citep[e.g.,][]{Dijkstra_and_Loeb2008,Chang2017}.

If the AGN is embedded in a dense ($n_{\rm H}>10^{5}~\mathrm{cm}^{-3}$) and compact ($\lesssim 200\pc$) neutral hydrogen reservoir, scattering and dust extinction can help reproduce the V-shaped SED typical of LRDs, in the presence of young PopIII stars. This scenario is in general consistent with the inferred densities in the nuclei of LRDs. By analysing the absorption of Balmer lines, a number of authors derive hydrogen densities of $>10^8~\mathrm{cm}^{-3}$ in the core of LRDs \citep[][]{Juodzbalis2024,Ji2025,Naidu2025,Taylor2025}. Moreover, \citet[][]{Inayoshi_and_Maiolino2025} showed that a Balmer break feature can originate in AGN spectra if the latter are embedded in extremely dense hydrogen cores, possibly featuring strong outflows. At these densities, assuming temperatures of $T>10^4~\mathrm{K}$, fragmentation in the nucleus can happen on mass-scales of $<10^3\Msun$, possibly favouring the formation massive PopIII stars. Finally, inflows form a quasi-isotropic and dense gas reservoir would enforce super-Eddington accretion onto newborn DCBHs \citep[e.g.,][]{Regan2019,Sassano2023,Lupi2024,Kao2025}, as expected for some LRDs \citep[e.g.,][]{Inayoshi2024,Madau2025}.

\section{Summary and conclusions}\label{sec:conclusions}
In this work, we explore the connection between Little Red Dots (LRDs) and direct-collapse black hole (DCBHs) that have recently formed. We employ preliminary runs from the upcoming \simname suite of cosmological simulations, where we implement a new accurate model for DCBH formation. Specifically, we form DBHs at centre of simulated haloes that are dense and pristine (gas-phase metallicity $<10^{-6}-10^{-4}~Z_\odot$), experience intense Lyman-Werner radiation and mass-inflow rates, and have not formed stars yet. In the following, we summarize our findings and conclusions:

\begin{itemize}
    \item[$\left(i\right)$] The population of newly-formed DCBHs in our simulations swiftly declines at $z\lesssim 6$ (see Figure~\ref{fig:BH_abundance}), akin to the emergence of LRDs \citep[][]{Kocevski2024}. Overall, low-redshift haloes are less likely eligible for DCBH formation, due to their lower central gas densities and higher metallicity than high-redshift haloes. Furthermore, below $z\simeq 6$, haloes in our simulations experience significantly reduced mass-inflow rates (see Figure~\ref{fig:inflow}).\\
    
    \item[$\left(ii\right)$] DCBHs exhibit an active phase with $L_{\rm bol}>10^{43}~\mathrm{erg}\,\mathrm{s}^{-1}$ in the $\lesssim 200\Myr$ after their formation, when they accrete close to their Eddington limit. Less than $2\percent$ of our DCBHs can grow to $\gtrsim 3\times 10^7\Msun$ during their active phase.\\
    
    \item[$\left(iii\right)$] In the $200\Myr$ prior to the formation of a DCBH, haloes experience efficient inflows where neutral gas free-falls towards their centre, that is associated with the increase in the central neutral gas density and a decrease in the effective size of the gas reservoir (see Figure~\ref{fig:gas_evolution}). After the DCBH has formed, the first PopIII stars also form, the gas reservoir becomes more diffuse and inflows less efficient.\\
    
    \item[$\left(iv\right)$] Rayleigh scattering off neutral hydrogen with extremely high column densities ($N_{\rm H}\gtrsim 10^{25}~\mathrm{cm}^{-2}-10^{26}~\mathrm{cm}^{-2}$) around newly-formed DCBHs could account for the expected UV attenuation and weak X-ray emission (see Figure~\ref{fig:LRD_SED}).

\end{itemize}

The continuum emission from the first PopIII stars contributing to the SED of DCBH nurseries can be distinguished from that of more evolved galaxies with observations with the JWST medium-band NIRCam filters \citep[e.g.,][]{Fujimoto2025}. Furthermore, accurate modelling of the SED of young and actively accreting DCBHs \citep[see, e.g.,][]{Natarajan2017,Natarajan2024,Inayoshi2022,Nakajima_and_Maiolino2022} will be crucial to potentially distinguish the DCBH nursery systems studied in this work among the population of LRDs.

DCBH nurseries arise in the 100 Myr after the formation of DCBHs, characterised by a dust-free, extremely dense, and compact neutral hydrogen envelope, surrounding a newborn DCBH, that accretes close to (or possibly exceeding) its Eddington limit. Our results suggest that a fraction of observed LRDs at $z>4$ could be associated with such DCBH nurseries \citep[as similarly proposed by][]{Inayoshi2025}, or an even earlier stages of DCBH formation \citep[][]{Begelman_and_Dexter2025}. However, LRDs may not trace an homogeneous population and it is currently hard to assess to what extent newly-formed DCBHs can contribute to the population of LRDs. With larger and higher-resolution simulations (\simname; Cenci et al., in preparation) we will be able to better probe the physical state of the nuclei of DCBH hosts and understand whether or not these system could correspond to the observed population of LRDs.

\section*{Acknowledgements}
The authors thank the anonymous referee for insightful comments that helped to improve the paper. EC thanks M. Volonteri for helpful comments and P. R. Capelo for inspiring discussion and support. EC thanks everyone at Ecogia, Department of Astronomy of UNIGE, for helpful discussion. MH and EC acknowledge support from the Swiss SNSF Starting Grant (grant no. 218032). The computations were performed at the University of Geneva, using \textsc{yggdrasil} HPC service, and at the Department of Astronomy of UNIGE using \textsc{bonsai} HPC service. All plots were created with the \textsc{matplotlib} library for visualization with Python \citep{Hunter2007}. 

\section*{Data Availability}
The data supporting the plots within this article are available on reasonable request to the corresponding author.



\bibliographystyle{mnras}
\bibliography{main}


\appendix
\section{Evolution of physical properties of haloes}\label{app:haloes}
\begin{figure*}
    \centering
    \includegraphics[width=\hsize]{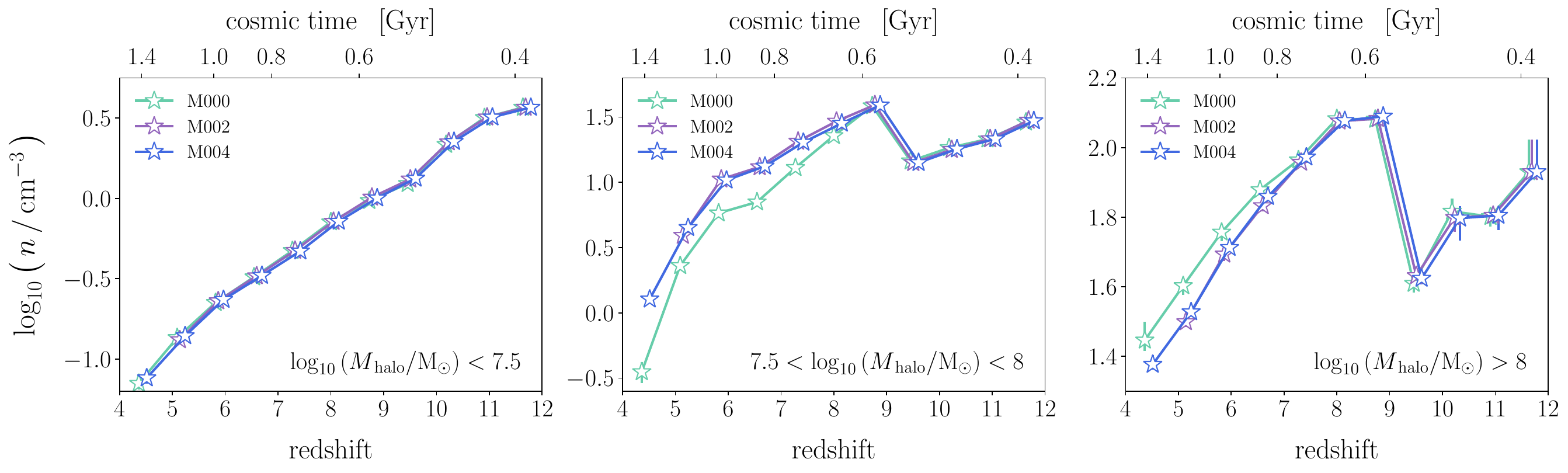}
    \caption{Average central (peak) hydrogen gas number density ($n$) in all haloes in our simulations, as a function of redshift, in different simulations, in three different halo mass bins. Data points are the median values with bootstrapped (16-th to 84-th percentiles) errorbars. In general, the median $\bar{n}_{\rm H}$ of haloes within a given mass bin decreases at lower redshifts. Haloes with central densities below $1~\mathrm{cm}^{-3}$ are not eligible for DCBH formation in our simulations. From the lowest to the highest mass bins, about $10$, $80$, and $97\percent$ of haloes still have peak densities compatible with DCBH formation at $z\simeq 6$.}
    \label{fig:evol_nH}
\end{figure*}

\begin{figure*}
    \centering
    \includegraphics[width=\hsize]{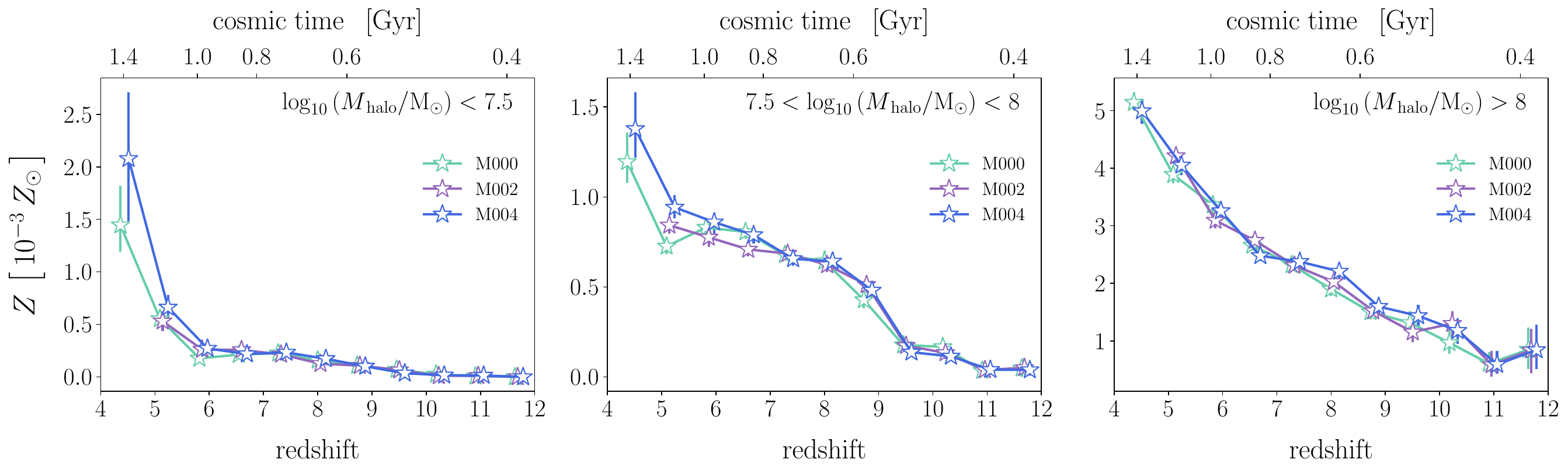}
    \caption{Gas-phase metallicity ($Z$) in all haloes in our simulations, as a function of redshift, in different simulations, in three different halo mass bins. Data points are the mean values with bootstrapped (16-th to 84-th percentiles) errorbars. On average, the metallicity of haloes of all masses increases at lower redshifts. Haloes with metallicities $Z>10^{-4}~Z_\odot$ in M000 and M002, and haloes with $Z>10^{-6}~Z_\odot$ in M004 are not eligible for DCBH formation. Despite the larger mean metallicity at lower redshift, the majority ($\gtrsim 85-90\percent$, depending on the simulation) of haloes with masses $<10^8\Msun$ at $z\simeq6$ still have metallicities below the critical threshold, while only about $50\percent$ of more massive haloes do.}
    \label{fig:evol_Z}
\end{figure*}

\NEW{In this section, we show the evolution of physical properties of haloes in the simulations used in this work, to complement the results shown in Figure~\ref{fig:inflow}. Specifically, we show the average evolution of the average central gas density (Figure~\ref{fig:evol_nH}) and gas-phase metallicity (Figure~\ref{fig:evol_Z}) within the haloes in three different mass bins. On average, low-redshift haloes have both lower central gas densities and higher metallicities. Our prescription for the formation of direct-collapse black holes (DCBHs) is based on both density and metallicity, as well as mass-inflow rates and other local halo properties (see Section~\ref{sec:DCBH_model_methods}). Therefore, the efficiency of DCBH formation over cosmic time is generally affected by the evolution of these quantities, with higher metallicities and lower gas densities and inflow rates contributing in hindering DCBH formation. In M000 and M002, we do not allow for DCBH formation if haloes have gas-phase metallicities $Z>10^{-4}~Z_\odot$, while in M004 we do not for DCBHs in haloes with $Z>10^{-6}~Z_\odot$. Furthermore, the central gas density threshold below which a halo is not eligible for DCBH formation is fixed at $1~\mathrm{cm}^{-3}$ in all the simulations presented here. This naturally results in a lower fraction of eligible haloes for DCBH formation according to our density and metallicity criteria at lower redshift and especially in simulation M004. Despite the mean metallicity of haloes becoming significantly larger than the maximum metallicity thresholds for DCBH formation at lower redshifts, most haloes still satisfy the metallicity criterion for DCBH formation. Specifically, the majority (about $\gtrsim 85-90\percent$, depending on the simulation) of haloes with total masses $M_{\rm halo}<10^8\Msun$ at $z\simeq6$ have metallicities below the critical threshold, while only about $50\percent$ of more massive ($M_{\rm halo}>10^8\Msun$) haloes do. Based on the minimum density criterion, about $10$, $80$, and $97\percent$ of haloes are still eligible for DCBH formation at $z\simeq 6$, from the lowest to the highest mass bins, respectively.}

%

\bsp	
\label{lastpage}
\end{document}